\documentclass[aps,prl,reprint,nofootinbib,superscriptaddress,secnumarabic,floatfix,longbibliography]{revtex4-2}
\usepackage[T1]{fontenc}
\usepackage[utf8]{inputenc}
\usepackage{lmodern}
\usepackage{microtype}

\usepackage{amsmath,amssymb,amsfonts,mathtools}
\usepackage{amsthm}
\usepackage{bm}
\usepackage{physics}
\usepackage{slashed}

\usepackage{graphicx}
\usepackage{xcolor}
\usepackage{hyperref}
\hypersetup{
  colorlinks=true,
  linkcolor=blue!60!black,
  citecolor=blue!60!black,
  urlcolor=blue!60!black
}

\theoremstyle{definition}

\makeatletter
\renewcommand\paragraph{\@startsection{paragraph}{4}{\z@}%
  {1.25ex \@plus 0.5ex \@minus 0.2ex}%
  {-0.8em}%
  {\normalfont\normalsize\bfseries}}
\makeatother

\begin{document}

\title{Dynamical paths in high-dimensional landscapes via optimal control theory}

\author{Thomas Louis-Sarrola}
\email{thomas.louis-sarrola@universite-paris-saclay.fr}
\affiliation{Université Paris-Saclay, CNRS, LPTMS, 91405 Orsay, France}

\author{Valentina Ros}
\affiliation{Université Paris-Saclay, CNRS, LPTMS, 91405 Orsay, France}

\author{Pierfrancesco Urbani}
\affiliation{Université Paris-Saclay, CNRS, CEA, Institut de physique théorique, 91191 Gif-sur-Yvette, France}

\date{\today}

\begin{abstract}
  Rare transitions in high-dimensional landscapes are often studied by conditioning them to occur within a finite horizon $\hat T$. Using optimal control, we solve this problem exactly for a quadratic random energy landscape on the  sphere, at any dimension $N$. We show that when the landscape is non-marginal, the paths converge to equilibrium instantons (time-reversal-symmetric) exponentially fast in $\hat T$. In the marginal case, the limits $N\to\infty$ and $\hat T\to\infty$ do not commute: mean-field instantons are different from equilibrium ones for any $\hat T$.
  On horizons $\hat T\sim N^{2/3}$, mean-field and equilibrium instantons are connected by a non-self-averaging crossover, which we solve exactly.
\end{abstract}

\maketitle
How stochastic dynamics carries a system between distinct regions of configuration space is a fundamental question, made especially challenging when that space is high-dimensional~\cite{laio2002escaping,  breebaart2026understanding, ray2023deep, tian2020exploring, mauri2023transition}. In large dimension, dynamics can often be characterized through observables that concentrate, whose distribution over the trajectory ensemble collapses onto typical values obeying a closed set of equations. These Dynamical Mean-Field Theory (DMFT) equations have been used to describe high-dimensional inference and optimization, population dynamics, biological and artificial learning \cite{mannelli2019passed, kamali2023stochastic, Folena2020Rithinking, fournier2023statistical, montanari2026dynamical}. This description, however, does not resolve individual paths, typical or not, and only accounts for dynamics at short times, that is, times of order one in the dimension.\\
This limitation becomes significant in the presence of trapping, when dynamics is confined by collective energetic or entropic barriers that grow with dimension, and escape occurs only through exponentially rare trajectories. Such barrier-crossing transitions are suppressed in the infinite-dimensional limit, and are thus outside the observation window of DMFT. Yet at large but finite dimension they control the long-time dynamics. How does a high-dimensional stochastic system realize them?\\
Addressing this question for complex high-dimensional systems requires studying stochastic dynamics at exponentially large times in dimension~\cite{franz2007dynamics}. This is considerably challenging, and has been carried out successfully only for dynamics on simplified, correlation-free random energy landscapes~\cite{benarousUniversalityREMDynamics2008, gayrardConvergenceClockProcesses2016, baityjesiActivatedAgingDynamics2018, starioloBarriersTrappingTimes2020}, where to some extent dynamics is  understandable via the extreme-value statistics of the landscape. An alternative is to sample barrier-crossing trajectories directly via large-deviation theory, conditioning for such rare processes to occur at timescales of order one; this retains concentration and the DMFT formalism, at the price of losing causality in the resulting dynamical equations~\cite{rizzoPathIntegralApproach2021a, charbonneauRareTrajectoriesPrototypical2026}. 
Large-deviation theory then offers a general framework for resolving transition paths explicitly, one that is potentially useful also from a control perspective, i.e., for manipulating the system to accelerate such transitions. The connection to optimal control is in fact quite deep~\cite{fleming2006stochastic, dupuis2011weak, jack2020ergodicity}: large-deviation calculations can be formulated as optimal-control problems, that become deterministic in the weak-stochasticity regime. There, a trajectory realizing a prescribed transition is governed by the cost of the noise realization driving it, and the most probable such trajectory solves the least-cost control problem steering the system between initial and final states. Optimal control theory then yields coupled forward–backward systems of equations for the optimal trajectory and its conjugate control field, whose solution gives an explicit realization of these rare transitions.\\
It remains an open question, however, to what extent these large-deviation trajectories capture the transitions that occur spontaneously over exponentially long times in high-dimensional systems. Here, we address this question by studying instantonic trajectories in quadratic landscapes, a benchmark problem that we solve exactly at arbitrary dimension and time horizon using the optimal-control formulation. We show that, in certain regimes, the optimal transition paths obtained in the infinite-dimensional limit differ qualitatively from those describing spontaneous transitions at long times in finite dimensions, demonstrating that the large-dimension and long-time limits need not commute.

\paragraph*{Rare paths via optimal control.} For concreteness, we consider Langevin dynamics in an energy landscape $ H({\bm \sigma})$ defined on  $ S^{N-1}(\sqrt N)= \left\{ {\bm \sigma}: \|{\bm \sigma}\|^2=N \right\} \in \mathbb{R}^N$,
\begin{align}
  \dot{\bm \sigma}(t)&=-\mu(t){\bm \sigma}(t)-\nabla H({\bm \sigma}(t))+{\bm \eta}(t).
  \label{eq:langevin-main}
\end{align}

Here $\mu(t)$ enforces the spherical constraint and ${\bm \eta}(t)$ is centred Gaussian noise with  $ \langle\eta_i(t)\eta_j(t')\rangle
  =2T\delta_{ij}\delta(t-t')$. We focus on the weak-noise limit $T \to 0$. 
Our goal is to characterize paths that achieve a prescribed degree of decorrelation from the initial condition ${\bm \sigma}(0)\equiv {\bm \tau}$ over a  finite time horizon $\hat T$; on the sphere, this can be done enforcing a value of the overlap
\begin{equation}\label{eq:Overlap}
 C(\hat T,0) =\frac1N{\bm \sigma}(\hat T)\cdot{\bm \tau} =q.
\end{equation}
Trajectories realizing this amount of decorrelation occur with probability
\begin{align}\label{eq:DPF}
  Z_{\bm \tau}(q, \hat T)
  &=\left\langle
  \delta\!\left(C(\hat T,0)-q\right)
  \,\middle|\,{\bm \sigma}(0)={\bm \tau}
  \right\rangle,
\end{align}
where the average is over realizations of the stochastic dynamics \eqref{eq:langevin-main}.
If decorrelation requires crossing a barrier, we expect this probability to decay in $N$ with a rate function \footnote{This rate function is directly connected to the dynamical Franz-Parisi potential studied in \cite{charbonneauRareTrajectoriesPrototypical2026} in the context of barrier-crossing events in disordered energy landscapes. That  is obtained averaging \eqref{eq:FinitePot} over the initial conditions drawn from a thermal measure at some inverse temperature $\beta_0$,  and subsequently taking the average $\overline{\cdot}$  over the randomness in $H$.
 }
\begin{equation}\label{eq:FinitePot}
  v_ {\bm \tau}(q, \hat T)= -\frac 1 N \log   Z_ {\bm \tau}(q, \hat T). 
  \end{equation}
 
 \begin{figure}[t]
  \centering
  \includegraphics[width=\columnwidth]{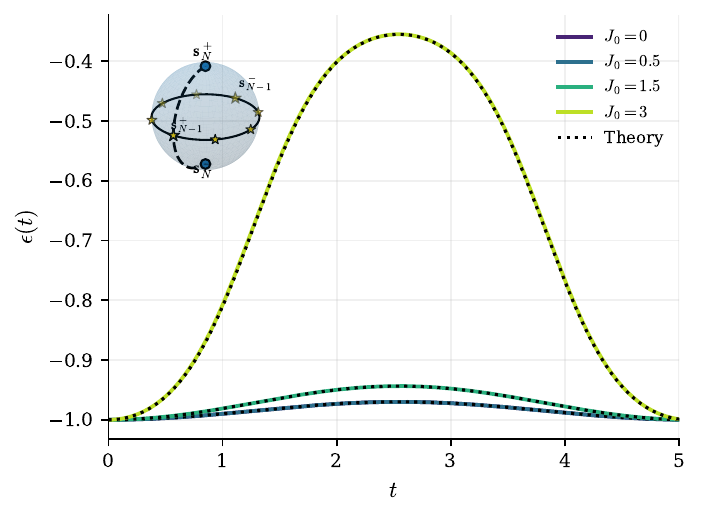}
  \caption{Energy-density profiles for different values of $J_0$.
    Solid lines are finite-$N$ forward--backward numerical solutions with
    $N=500$, $\hat T=5$, $\tilde\nu=10$, and $N_{\hat T}=500$ time steps.
    Dotted lines show the exact solution, obtained from the reduced dynamics~\eqref{eq:Hamilton}.
    An additive constant is applied to each energy profile so that $\epsilon(0)=-1$. Inset: Sketch of the optimal path in configuration space.}
  \label{fig:beta_infty_numerics}
\end{figure}

We are interested in characterizing both the probability associated with the dynamical paths, controlled by \eqref{eq:FinitePot}, and the mechanism by which such events are realized, i.e., the explicit form of the corresponding trajectories. Both these aspects are directly addressed by formulating the problem in the language of optimal control theory. The optimal control approach applied to this problem leads to a particularly simple prescription in the $T\to 0$ limit:  it requires to solve a \emph{deterministic} Forward-Backward (FB) system of equations for the $N$-dimensional trajectory ${\bm \sigma}$ and the associated control field ${\bm \pi}$,
\begin{equation}\label{eq:sigma-pi}
\begin{split}
  \dot {\bm \sigma}(t)
  &=
  -\mu(t) {\bm \sigma}(t)
  -
  \nabla H( {\bm \sigma}(t))
  +
  {\bm \pi}(t) 
  \\
  \dot {{\bm \pi}}(t)
  &=
  \left[
    \mu(t)
    +
    \nabla^2 H( {\bm \sigma}(t))
  \right]{\bm \pi}(t)
  +z(t){\bm \sigma}(t)
  \end{split}
\end{equation} 
where $\mu$ and $z$ are fixed by imposing  the spherical constraint on ${\bm \sigma}$ and the requirement that the control field remain tangent to the sphere, ${\bm \sigma}\cdot{\bm \pi}=0$, at all times. This construction is generic for spherical models: it ensures that no control effort is wasted along the radial direction. The boundary conditions are imposed at the initial and final times, respectively: $ {\bm  \sigma}(0) = {\bm \tau}$, and ${\bm \pi}(\hat{T}) = -\tilde{\nu} P_{{\bm \sigma}(\hat{T})}[{\bm \tau}]$, where $P_{{\bm \sigma}}[\cdot]$ projects onto the tangent space of the sphere at ${\bm \sigma}$,  and  $\tilde{\nu}$ is a real parameter labelling the family of solutions. The trajectories $({\bm \sigma}^*, {\bm \pi}^*)$ solving \eqref{eq:sigma-pi} are optimal in the following sense: the control ${\bm \pi}^*$ realizes the infimum of the cost functional 
\begin{equation}\label{eq:Cost}
\mathcal{C}_{\hat T, \tilde \nu}[{\bm \pi}; {\bm \tau} ]= \frac{1}{2 N} \int_{0}^{\hat T} dt || {\bm \pi}(t)||^2 +  \frac{\tilde{\nu}}{N}{\bm \sigma}^{\bm \pi}(\hat T)\cdot{\bm \tau} 
\end{equation}
subject to the controlled dynamics ${\bm \sigma}^{\bm \pi}$ defined by Eq.~\eqref{eq:sigma-pi}. The parameter $\tilde{\nu}$  labelling the solutions is conjugate to the overlap in Eq.~\eqref{eq:Overlap}, the relation between the two being obtained by differentiating the function
\begin{equation}\label{eq:LT}
\tilde{\psi}_ {\bm \tau}( \tilde \nu, \hat T):= - \mathcal{C}_{\hat T, \tilde \nu}[{\bm \pi}^*; {\bm \tau} ], \quad - \frac{\partial\tilde{\psi}_ {\bm \tau}( \tilde \nu, \hat T)}{\partial \tilde{\nu}}=q.
\end{equation}
In other words, the control problem is first solved as a function of $\tilde \nu$, whose value parametrizes the family of attainable final configurations ${\bm \sigma}(\hat{T})$; one then fixes $\tilde\nu\to \tilde \nu(q)$
as the value of $\tilde \nu$ that yields the target final overlap $q$, via the derivative relation above. 
Solving \eqref{eq:sigma-pi} allows one to characterize the rare dynamical processes: ${\bm \sigma}^*$ is the trajectory that realizes optimally the decorrelation in finite time  $\hat T$, while ${\bm \pi}^*$ quantifies how the system is optimally steered toward the final configuration, and it also controls the probability cost of the trajectory \eqref{eq:FinitePot}, 
\begin{equation}\label{eq:RescaledPot}
\tilde{v}_{\bm \tau}(q, \hat T):=\lim_{T \to 0} 2 T \, v_{\bm \tau}(q, \hat T)= \frac{1}{2 N} \int_{0}^{\hat T} dt || {\bm \pi}^*(t)||^2.
\end{equation}

Underlying this optimal control formulation is, in fact, a large-deviation calculation: in the weak-noise limit, averages over stochastic trajectories like \eqref{eq:DPF} are dominated by a single, optimal noise realization, giving rise to the \emph{instantonic} trajectory $\bm\sigma^*$: this is the dominant saddle-point path controlling the exponentially small probability of the corresponding rare event \cite{freidlinRandomPerturbationsDynamical2012}. The optimal control field describes precisely this noise trajectory; $\tilde{\nu}$ is then a tilt parameter, that controls how the ensemble of trajectories has to be reweighted to bias it towards these atypical ones. There are therefore two equivalent views of an atypical path: a rare realization of the noise, or a controlled path driven by the optimal field ${\bm \pi}^*$. We elucidate this equivalence in the End Matter.

\paragraph*{Solving by causal sweeps.} Optimal control has been used in the stochastic-processes literature to characterize rare paths~\cite{smelyanskiy1997optimal, ControlledMarkovProcesses2006, chernyak2014stochastic, chetrite2015variational}, but its potential in high dimensions~\cite{urbaniDisorderedHighdimensionalOptimal2021a} is underexploited. Rare paths in this setting have been studied via large deviations combined with DMFT: conditioning on both initial and final configurations, and solving for the typical $N \to \infty$ trajectories matching these boundaries. This \emph{conditional DMFT} yields equations with a non-causal structure~\cite{rizzoPathIntegralApproach2021a, charbonneauRareTrajectoriesPrototypical2026}, since the dynamics must be consistent with a boundary condition set in the future. Similarly, the boundary condition for the control field in \eqref{eq:sigma-pi} is at the horizon time $\hat T$. At finite $N$, though, the control formulation splits the problem into an initial-value problem for ${\bm \sigma}$ at given ${\bm \pi}$ and a terminal-value problem for ${\bm \pi}$  at given ${\bm \sigma}$, integrated forward and backward in time. Each forward/backward sweep is causal, and alternating them gives an iterative scheme~\cite{chernykh2001large}. Since each sweep is an ordinary causal dynamics, its large-$N$ limit is a standard DMFT, suggesting an alternative to conditional DMFT (see Discussion). In simple cases, this iteration can even be solved exactly at arbitrary $N$, giving analytic access to rare paths at arbitrary dimension. We devote the rest of the paper to showing this.

\paragraph*{Quadratic Gaussian landscapes.} We consider random Gaussian landscapes of the form 
\begin{equation}  \label{eq:ssk-main}
  H({\bm \sigma})=-\frac12{\bm \sigma}^{\mathsf T}J{\bm \sigma},
  \qquad
  J=W+\frac{J_0}{N}\mathbf1\mathbf1^{\mathsf T},
\end{equation}
where $W$ is a GOE matrix with centred entries and off-diagonal variance $1/N$: this is a soft-spin version of the Sherrington–Kirkpatrick model \cite{kosterlitzSphericalModelSpinGlass1976}. We assume $J {\bf v}_\alpha=\lambda_\alpha {\bf v}_\alpha$ with $ \lambda_1\leq \lambda_2 \leq  \cdots \leq \lambda_N$. Owing to its quadratic nature, the landscape is fully characterized by the spectrum of $J$: there are $2N$ stationary points ${\bf s}_\alpha^{\pm}=\pm \sqrt{N}{\bf v}_\alpha$, with energy density $\epsilon_\alpha= H({\bf s}_\alpha^\pm)/N = -\lambda_\alpha/2$ and index  $N-\alpha$.  Thus the landscape admits only two degenerate stable points, the minima ${\bf s}_N^{\pm}$. Since the eigenvectors are orthogonal, the minima may be viewed as the poles of $S^{N-1}$, while all saddles ${\bf s}_{\alpha \neq 1, N}^{\pm}$  lie on the equator (see the inset of Fig.~\ref{fig:beta_infty_numerics}). Escaping a minimum requires crossing a saddle, with associated energy-density barrier $\Delta \epsilon_\alpha = (\lambda_N -\lambda_\alpha)/2$. In particular, the lowest barrier in energy density is $\Delta \epsilon_{N-1} = \delta^{(N)}/2$, where 
\begin{equation}
  \delta^{(N)}:=\lambda_N-\lambda_{N-1} \sim   \delta^{(\infty)}+ N^{-\zeta} g_N
  \label{eq:gap-main}
\end{equation}
 is the spectral gap of $J$, the superscript denotes the dependence on the size $N$ of $J$ and $g_N$ is an $O(1)$ random variable. The statistics of $\delta^{(N)}$ undergo a transition at  $J_0=1$ inherited by the BBP transition of $\lambda_N$  \cite{edwardsEigenvalueSpectrumLarge1976}: when $N\to\infty$,  for $J_0<1$ the gap vanishes, $\delta^{(\infty)}=0$, while for $J_0>1$ it converges to $\delta^{(\infty)} = J_0+J_0^{-1}-2$. The transition widens into a critical window of size $N^{-1/3}$ at large but finite $N$ \cite{bloemendal2013limits}, with $\zeta =2/3$ in the subcritical regime and $\zeta =1/2$ in the supercritical regime.  
 Relaxational dynamics of this model has been studied through DMFT in \cite{cugliandoloFullDynamicalSolution1995}, and beyond mean field in~\cite{fyodorovLargeTimeZero2015, barbier2021finite, pimenta2023finite}. Activated dynamics (trajectories transitioning between the degenerate minima) have been studied in \cite{lopatinInstantonsLangevinDynamics1999} under the assumption that instantons are \emph{equilibrium}, i.e., they satisfy detailed balance. Using optimal control, we now solve the escape problem at arbitrary $N$ and horizon $\hat{T}$. 

\begin{figure}[t]
  \centering
  \includegraphics[width=0.48\columnwidth]{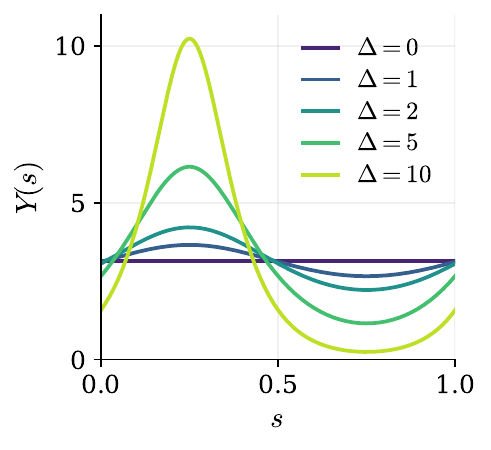}
  \hfill
  \includegraphics[width=0.48\columnwidth]{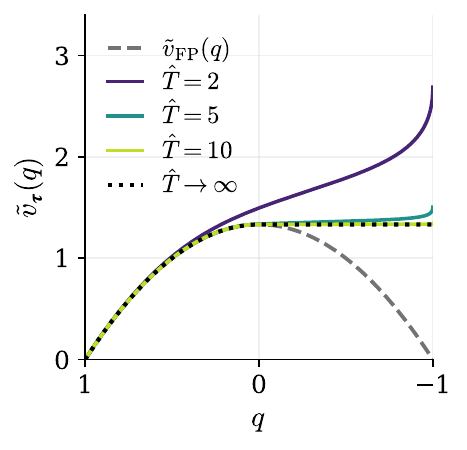}
  \caption{\textit{Left:} Optimal-control amplitude $Y(s)$ for the
    target overlap $q=-1$ and $\Delta=0,1,2,5,10$.
    \textit{Right:} Dynamical rate function $\tilde v_{\bm\tau}(q,\hat T)$
    for $J_0=3$ and $\hat T=2,5,10$, compared with the
    Franz--Parisi potential $\tilde v_{\rm FP}(q)$ and the
    $\hat T\to\infty$ dynamical limit.}
  \label{fig:validation}
\end{figure}

\paragraph{Exact solution of the FB system.} 
We characterize paths escaping one minimum, ${\bm \tau} ={\bf s}_N^+= \sqrt{N}{\bf v}_N$, with a decorrelation controlled by $q$. Realizing such a transition optimally involves two problems: (i) the geometry: which route on the sphere is chosen to move away from the pole, and (ii) the kinetics: how that route is traversed in time. For arbitrary $N$, the FB system can be solved by iteration; Fig.~\ref{fig:beta_infty_numerics} shows the resulting energy-density profile $\epsilon(t)= H({\bm \sigma}(t))/N$ for $q=-1$ (trajectories interpolating between the two minima). Inspecting the iterative solutions suggests a simple structure. Decomposing the FB process into the modes $\sigma_\alpha= {\bm \sigma}\cdot {\bf v}_\alpha$ one finds an exact solution (dotted lines in Fig.~\ref{fig:beta_infty_numerics}), derived in the Supplemental Material (SM) and summarized here. \\
\indent \emph{The geometry.} Any trajectory initialized in a subspace spanned by ${\bm \tau} = {\bf s}^+_N$ and other eigenvectors ${\bf s}^{\pm}_\alpha$ remains in that subspace. It turns out that the system always selects a fixed direction to decorrelate, corresponding to $\alpha=N-1$: it moves towards the lowest saddle. This is true not only for $q<0$, where trajectories must cross the equator passing through a saddle, but also for $q>0$. This is likely because in this model the same direction is also that of softest curvature around the minimum, making it optimal for moving away. The escape path can then be parametrized (up to the degeneracy $\theta \to -\theta$) as
\begin{equation}  \label{eq:geodesic-main}
\begin{split}
  {\bm \sigma}(t)&=\cos\theta(t)\,{\bf s}^+_N+\sin\theta(t)\,{\bf s}_{N-1}^+,  
  \\
  {\bm \pi}(t)&=y(t)\left[-\sin\theta(t)\,{\bf s}^+_N+\cos\theta(t)\,{\bf s}_{N-1}^+\right],
  \end{split}
\end{equation}
with $y(t)$ and $\theta(t)$ to be determined: the high-dimensional escape problem is thus reduced to a one-dimensional one for an angle and its associated velocity.

\indent \emph{The kinetics.} $y(t)$ and $\theta(t)$ depend on the parameters $\tilde{\nu}$, $\hat{T}$ and $\delta^{(N)}$. Substituting \eqref{eq:geodesic-main} into the FB system and introducing the scaling parameters $s= t/{\hat{T}}$, $ \kappa= \tilde{\nu}\hat{T}$ and $\Delta \equiv \Delta^{(N)}=\delta^{(N)}\hat{T}$, one finds a reduction to the scaling functions $\theta(t; \tilde{\nu}, \hat{T}, \delta^{(N)}) \to  \Theta\left(s;  \kappa, \Delta \right)$ and $y(t; \tilde{\nu}, \hat{T}, \delta^{(N)})\to   Y\left(s;  \kappa, \Delta\right)/\hat{T}$. 
These are conjugate variables following conservative dynamics with Hamiltonian
\begin{equation}\label{eq:Hamilton}
\mathcal{H}=\frac{N}{2}\left([Y\left(s;  \kappa, \Delta\right)]^2- \Delta Y\left(s;  \kappa, \Delta\right)\sin [2 \Theta\left(s;  \kappa, \Delta\right)]\right).
\end{equation}
The equations of motion are $\dot Y = \Delta  Y\cos(2\Theta)$, $\dot \Theta = Y-\frac{\Delta}{2}\sin(2\Theta)$, with boundary conditions $\Theta(0) = 0$, $Y(1) = \kappa \sin[\Theta(1)]$. Tuning  $\kappa$ one gets the optimal control for various $q=\cos \Theta(1)$. The Hamiltonian form is built into the optimal-control formalism, and energy conservation yields explicit solutions for the kinetics, given in the End Matter. Dependence on the randomness and on $N$ enters only through the rescaled spectral parameter $\Delta^{(N)}$.

\paragraph{Mean-Field (MF) instantons.}
Mean-field ($N \to \infty$) instantons are the processes we expect to recover through approaches like conditional DMFT. In MF, the optimal control field exhibits a transition: for $J_0\leq1$ the rescaled gap $\Delta^{(\infty)}=0$ and \eqref{eq:Hamilton} reduces to the kinetic term alone, so $Y$ is constant in $s$, implying $||{\bm \pi}^*(t)||= \sqrt{N}\arccos(q)/\hat T$. The optimal field drives the system to the final configuration at uniform velocity, set by $q, \hat{T}$. We call these \emph{marginal} MF instantons, since in this regime the curvature at ${\bm \tau}$ vanishes as the spectral gap closes. They describe genuine escape processes, though: the energy barrier scales as $N^{1/3}$ and still diverges, even though the barrier in energy \emph{density} vanishes.

For $J_0 >1$, $\Delta^{(\infty)}>0$ and $Y= \dot \Theta +\Delta^{(\infty)} \sin(2\Theta)/2$ acquires a non-trivial $s$-dependence: it has to provide the velocity and (for $\Theta<\pi/2$) counteract the landscape force $\Delta^{(\infty)} \sin(2\Theta)/2$ which pulls the system back to ${\bm \tau}$. Plots of the resulting optimal control are given in Fig.~\ref{fig:validation} \emph{Left} for $\Theta(1)=\pi$. $Y$ depends on the path only through the angle (see SM): it increases up to $\Theta=\pi/4$, where the landscape is steepest and holding still would be costly, decreases up to $\Theta=3\pi/4$, and increases again towards $\Theta=\pi$, where it supplies the velocity needed to reach the target in finite time. A trajectory ending at $\Theta(1)<\pi$ only visits the part of this profile with $\Theta\le\Theta(1)$. As $\Delta^{(\infty)}$ increases, the control becomes localized around the time $s^*$ at which the path crosses $\Theta=\pi/4$, with shape $Y(s)\simeq\Delta^{(\infty)}\,\mathrm{sech}[\Delta^{(\infty)}(s-s^*)]$, truncated at $s=1$ if $\Theta(1)<\pi/4$. Correspondingly, since $Y=2\dot\Theta$ on the uphill branch, the trajectory takes the form of a kink, $\tan\Theta(s)=e^{\Delta^{(\infty)}(s-s^*)}$, steepest at $\Theta=\pi/4$, separated by plateaux at the stationary points, where the system lingers at no cost. The localization is complete as $\Delta^{(\infty)}\to\infty$ ($ \hat T\to \infty$). In this limit one expects to recover the \emph{equilibrium} instantons of \cite{lopatinInstantonsLangevinDynamics1999}, i.e., the escape processes that occur spontaneously at exponentially large times. We verify that this is the case: in this limit the uphill part of the path (${\bm \sigma}\cdot{\bm \tau}>0$) is the time reversal of the gradient-flow relaxation towards ${\bm \tau}$, and the rate \eqref{eq:RescaledPot} converges exponentially fast in $\Delta^{(\infty)}$ to
\begin{equation}\label{eq:FranzParisip2}
  \tilde v_{\bm \tau}(q,\hat T) \stackrel{\hat T\to\infty}{\longrightarrow}
  \begin{cases}
    \tilde v_{\rm FP}(q)=\delta^{(\infty)}(1-q^2), & 0\le q\le 1,  \\
    \delta^{(\infty)},                             & -1\le q\le 0,
  \end{cases}
\end{equation}
as shown in Fig.~\ref{fig:validation} \emph{Right}. Here $\tilde v_{\rm FP}(q)$ is the zero-temperature Franz-Parisi potential \cite{franzRecipesMetastableStates1995a}. Both facts reflect equilibrium (they follow from detailed balance), as discussed in the End Matter.

\begin{figure}[t]
  \centering
  \includegraphics[width=\columnwidth]{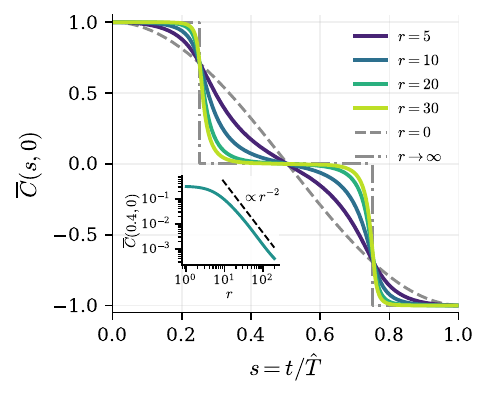}
  \caption{Crossover of the disorder-averaged correlation for $N=100$,
    $\kappa=100$, and $r=\hat T/N^{2/3}$. The dashed curve is the exact
    $r=0$ solution. The dash-dotted curve shows the sharp $r\to\infty$
    kink profile. Each finite-$r$ curve is averaged over 20 disorder realizations. Inset: $\overline{C}(s=0.4,0)$ versus $r$, for $\kappa=100$, averaged over $10^4$ gaps $g_N$ at $N=10^3$.}
  \label{fig:crossover}
\end{figure}

\paragraph{Beyond Mean-Field.}
Marginal MF instantons are insensitive to the energy profile along the path, crossing it at constant speed. They become infinitely slow as $\hat T \to \infty$. To get instantons that adjust optimally to the landscape, one needs time horizons $\hat{T}$ scaling with $N$. The dependence of the exact solution on $\Delta^{(N)}$ indicates that the natural timescale needed to traverse these paths is $\hat{T} \sim [\delta^{(N)}]^{-1}$: while for $J_0>1$ this remains $O(1)$ allowing to get equilibrium instantons within MF, for $J_0<1$ it diverges. The exact solution allows us to study the relevant crossover regime $\hat{T} \sim N^{2/3}$. At these large horizon times, no concentration occurs, and the solutions depend on the randomness through $\Delta^{(N)}= r g_N$, where $r \equiv \hat T N^{-2/3}$ and $g_N \sim O(1)$ is the rescaled gap. 

The parameter $r$ interpolates between MF instantons ($r \to 0$) and equilibrium ones ($r \to \infty$). Fig.~\ref{fig:crossover} shows the disorder-averaged correlation profile for increasing values of $r$. As $r$ grows, plateaux develop around the stationary points, consistent with the equilibrium limit. The asymptotic profile coincides with the one reached by the MF correlation for $J_0 > 1$ as $\Delta^{(\infty)} \to \infty$. However, while the MF convergence is exponentially fast, here it is only algebraic, scaling as $1/r^2$. This slower convergence reflects the non-self-averaging nature of this dynamical regime: for a single realization of the disorder the convergence is exponential, but the average is dominated by rare realizations in which the gap $g_N$ is atypically small.

\paragraph{Discussion.}
The optimal-control formulation of large deviations provides a useful framework for rare dynamical paths in high dimensions. For the escape problem in spherical quadratic random landscapes (in the weak-noise limit), it yields an exact solution for the dominant path and its probability cost at arbitrary system size and time horizon. Our central finding is that the limits $N \to \infty$ and $\hat T \to \infty$ need not commute. In non-marginal landscapes they do: as $\hat T$ grows, the optimal paths converge, exponentially fast in units of the relaxation time $1/\delta^{(N)}$, to the equilibrium instantons that Langevin dynamics realizes spontaneously on exponentially large timescales, while at finite $\hat T$ they describe genuinely non-equilibrium escapes, with a nonzero control Hamiltonian. In marginal landscapes they do not: the mean-field instanton remains a non-equilibrium process for any $\hat T$, so the mean-field large-deviation description and the exponentially long-time dynamics capture different escape mechanisms. This may be relevant for rugged landscapes, where mean-field activated jumps are likewise not localized in time but spread over the whole horizon~\cite{rizzoPathIntegralApproach2021a}, and different from what one would expect from equilibrium~\cite{rosDynamicalInstantonsActivated2021a}. We characterize the resulting crossover analytically, on horizons $\hat T \sim N^{2/3}$: there the dynamics is non-self-averaging, and disorder-averaged observables approach their equilibrium form only algebraically, as $1/r^2$, owing to atypically small spectral gaps. This extends existing results on relaxational dynamics beyond mean field to activated processes.

Beyond this solvable benchmark, the optimal-control formulation applies to genuinely rugged landscapes, such as the spherical $p \geq 3$ model. There, exponentially many saddles and escape routes, whose statistics are known~\cite{ros2020distribution, ros2019complexity}, compete in shaping the optimal path, and the selected route may change abruptly with $q$ or $\hat T$, giving rise to dynamical transitions. The forward--backward structure also suggests a route to large $N$ that is alternative to conditional DMFT: deriving a DMFT for the iteration scheme itself, following~\cite{urbani2026theory}. Such a description remains causal, and its fixed point encodes the large-deviation path; this is the subject of ongoing work. Finally, extending the approach to finite temperature requires stochastic rather than deterministic optimal control.

\paragraph{Acknowledgments.} We dedicate this work to the memory of our friend and colleague Berislav Bu\v{c}a. We thank the Zentrum f\"ur interdisziplin\"are Forschung (ZiF) in Bielefeld for hospitality during the program \emph{The Lush World of Random Matrices}, and Gabriele Sicuro for discussions. VR acknowledges funding by the French National Research Agency under grant ANR-26-CE57-2109 (HiPACO). PU acknowledges funding by the French government under the France 2030 program (PhOM - Graduate School of Physics) with reference ANR-11-IDEX-0003.

\paragraph{AI disclosure.} All ideas, calculations and technical strategies were proposed and derived by the authors. AI tools were used at intermediate stages of the work to generate the code for the figures, to check simple steps in the calculations, and to improve the clarity of the writing.

\bibliography{biblio.bib}

\clearpage
\section*{End Matter: Large deviations through optimal control}
 Optimal control problems are formulated in terms of value functions, which encode the cost-to-go from a given state to the prescribed final configurations.
In our problem, this role is played by the value function 
\begin{equation}\label{eq:ValFunc}
    V(t,{\bf x};{\bm \tau}) = -\log \left\langle e^{-\nu N C(\hat{T},0)}\, \middle| \,
 {\bm \sigma}(t) = {\bf x}
 \right\rangle,
\end{equation}
defined for ${\bf x} \in S^{N-1}(\sqrt N)$. 
The average is over trajectories obeying \eqref{eq:langevin-main}, the exponential factor providing a tilt that re-weights the trajectories as a function of the overlap $C(\hat{T}, 0)$ in \eqref{eq:Overlap}. The value function \eqref{eq:ValFunc} evaluated at $t=0$ is, up to a sign, the logarithm of the generating function associated with the probability \eqref{eq:DPF}.

Using the Feynman--Kac formula~\cite{kacDistributionsCertainWiener1949}, one can show that this quantity solves the HJB equation~\cite{ControlledMarkovProcesses2006}:
\begin{equation}
  \partial_t  V
  -
  \left[
    \mu({\bf x}){\bf x}
    +
    \nabla H({\bf x})
    \right]\cdot \nabla_S  V
  +
  T\Delta_S  V
  -
  T\|\nabla_S  V\|^2
  =
  0,
  \label{eq:hjb_terminal}
\end{equation}
with terminal condition $V(\hat{T},{\bf x};{\bm \tau})=\nu {\bf x}\cdot{\bm \tau}$. Notice that in the class of models that we are considering, $V(t,{\bf x};{\bm \tau})$ is defined on an hypersphere (${\bf x} \in S^{N-1}(\sqrt N)$); therefore, $\nabla_S$ is the Riemannian gradient of the function (that is tangential to the sphere) and not the Euclidean one, and $\Delta_S$ is the Laplacian on the sphere.
Classical results in optimal-control theory show that the tilted dynamics for arbitrary noise strength can be written as a forward--backward stochastic differential equation:
\begin{equation}\label{eq:BF_Noise}
  \begin{split}
    \dot {\bm \sigma}(t)
     & =
    -\mu(t){\bm \sigma}(t)
    -
    \nabla H({\bm \sigma}(t))
             +
             {\bm \pi}(t)
             +
             {\bm \eta}(t),
    \\
    \dot {{\bm \pi}}(t)
     & =
    \left[
      \mu(t)
      +
      \nabla^2 H({\bm \sigma(t)})
      \right]{\bm \pi}(t)
    +z(t){\bm \sigma}(t) +\Gamma(t){\bm \eta}(t),
  \end{split}
\end{equation}
where we have defined
\begin{equation}\label{eq:DEPHJB}
  \begin{split}
    {\bm \pi}(t) & := -2T\, \nabla_S V(t,{\bm \sigma}(t); {\bm \tau}),   \\
    \Gamma(t)    & := 2T \, \nabla^2_S V(t,{\bm \sigma}(t); {\bm \tau}).
  \end{split}
\end{equation}
The boundary conditions are ${\bm \sigma}(0)={\bm \tau}$ and
${\bm \pi}(\hat{T})=-2 T \nu P_{{\bm \sigma}(\hat T)}[{\bm \tau}]$. The fact that $\nabla_S$ is the Riemannian gradient shows immediately that  ${\bm \sigma} \cdot{\bm \pi}=0$ at all times.

This approach holds for arbitrary strength of the stochasticity, controlled by the temperature $T$. However, for general $T$ the term $\Gamma(t)$ cannot be computed without solving the full HJB equation, which is extremely challenging in high dimensions. In the weak-noise limit $T\to 0$, the HJB drastically simplifies and the system of equations \eqref{eq:BF_Noise} reduces to the deterministic system considered in the main text, with $\tilde{\nu}= 2 T \nu$. The forward--backward system can have several solutions: the relevant saddle is the one that minimizes this cost functional in Eq. \eqref{eq:Cost}.

We now illustrate how the $T \to 0$  forward--backward system is recovered from a large deviation calculation. The starting point is the generating function associated to the probability \eqref{eq:DPF},
\begin{align}\label{eq:GenF}
  \hat{Z}_{\bm \tau}(\nu, \hat{T}) & =\left\langle e^{-\nu NC(\hat T,0)}
  \,\middle|\,{\bm \sigma}(0)={\bm \tau}\right\rangle,
\end{align}
with the corresponding rate function
\begin{equation}\label{eq:rateGen}
  \psi_{\bm \tau}(\nu, \hat{T})=\frac1N\log   \hat{Z}_{\bm \tau}(\nu, \hat{T}).
\end{equation}
The rate \eqref{eq:FinitePot} can be obtained from \eqref{eq:rateGen} applying a Legendre--Fenchel transform. Notice that the value function satisfies $V(0,{\bm \tau}; {\bm \tau})=-N \psi_{\bm \tau}(\nu, \hat{T})$.

The quantity $ \hat{Z}_{\bm \tau}(\nu, \hat{T})$ is the partition function of a tilted dynamics
that favors particular realizations of ${\bm \sigma}$, and it can be written as a path integral over the noise as~\cite{martinStatisticalDynamicsClassical1973}:
\begin{align}
  \hat{Z}_{\bm \tau}(\nu, \hat{T})
   & =
  \int \mathcal{D}{\bm \eta}
  \int_{{\bm \sigma}(0)={\bm \tau}}
  \mathcal{D}{\bm \sigma}\mathcal{D}{\hat{\bm \sigma}}
  \int\mathcal{D}{\mu}\mathcal{D}{z}
  \nonumber       \\
   & \quad\times
  \exp\left(
  -\frac{1}{4T}\int_0^{\hat{T}} \dd{t} \|{\bm \eta}(t)\|^2
  - \nu {\bm \sigma}(\hat{T}) \cdot {\bm \tau}
  \right)
  \nonumber       \\
   & \quad \times
  \exp\left(
  \int_0^{\hat{T}} \dd{t}\,
  i\hat{\bm \sigma}(t) \cdot
  \left[\dot{\bm \sigma}(t)+\mu(t){\bm \sigma}(t)\right.
    \right.
    \nonumber\\[-1mm]
    &\hspace{5.7em}\left.
    \left.{}+\nabla H({\bm \sigma}(t))-{\bm \eta}(t)\right]
  \right)
  \nonumber       \\
   & \quad \times
  \exp\left(
  \frac{1}{2T}\int_0^{\hat{T}} \dd{t}\,
  z(t) \left(\|{\bm \sigma}(t)\|^2- N\right)
  \right).
\end{align}

When  $T\to0$, the rate function is obtained through a saddle-point calculation. We introduce the rescaled variables ${\bm \pi}(t)=-2T\,i\hat{\bm \sigma}(t)$ and $\tilde{\nu}=2T\nu$. Variation with respect to ${\bm \eta}$ gives ${\bm \pi}(t) ={\bm  \eta}(t)$, showing that in the weak-stochasticity limit, the noise indeed plays the role of a control.
The remaining saddle-point equations reproduce the forward-backward system \eqref{eq:sigma-pi}.
Varying with respect to ${\bm \sigma}(\hat{T})$ gives the terminal condition for ${\bm \pi}$ also reported in the main text. The functions $z$ and $\mu$ enforce ${\bm \sigma}^2(t)=N$ and ${\bm \sigma}(t)\cdot{\bm \pi}(t)=0$, respectively, and their explicit form reads:
\begin{align}
  \mu(t) & = - \frac{{\bm \sigma}(t) \cdot \nabla H({\bm \sigma}(t))}{N},                                      \\
  z(t)   & = \frac{1}{N}\left[
                          \nabla H\cdot{\bm \pi}-{\bm \sigma}\cdot(\nabla^2H){\bm \pi}-\|{\bm \pi}\|^2\right].
\end{align}
Plugging these expressions back into the action, one gets the rescaled rate function
\begin{equation}
  \begin{split}
    \tilde{ \psi}_{\bm \tau}(\tilde{\nu}, \hat{T}) & = \lim_{T\to 0} 2T  \psi_{\bm \tau}\left(\frac{\tilde{\nu}}{2 T}, \hat{T}\right)                                         \\
                                                   & = - \frac{1}{2N}\int_0^{\hat{T}} \dd{t} \| {\bm \pi}(t)\|^2 - \frac{\tilde\nu}{N} {\bm \sigma}(\hat{T})\cdot {\bm \tau},
  \end{split}
\end{equation}
consistently with Eq. \eqref{eq:LT} in the main text.  The Legendre--Fenchel transform gives the relation between the conjugate parameters $\tilde{\nu}$ and $q$, as well as Eq. \eqref{eq:RescaledPot} in the main text.

\section*{Exact solution of the FB system}
The reduced functions $\Theta(s), Y(s)$ follow conservative dynamics with Hamiltonian \eqref{eq:Hamilton}. The solutions are parametrized in terms of the control Hamiltonian density $u=2 \mathcal{H}/N$, conserved in time. Evaluating this quantity at the final configuration $\Theta_f \equiv  \Theta(1)$ yields 
\begin{equation}
u=  \kappa \sin \Theta_f \left( \kappa \sin \Theta_f - \Delta \sin(2 \Theta_f) \right).
\end{equation}
The trajectory $\Theta(s)$ is parametrized by $u$ (and thus by the final condition $\Theta_f$) and given implicitly by
\begin{equation}\label{eq:TS}
    s=\int_0^{\Theta(s)} \frac{d\phi}{\sqrt{u+ \frac{\Delta^2}{4} \sin^2 2 \phi}}.
\end{equation}
For fixed $\kappa, \Delta$, the final condition 
 $\Theta_f$ is obtained solving the self-consistent equation obtained plugging $s \to 1$ and $\Theta(s) \to \Theta_f$ in \eqref{eq:TS}.
The amplitude of the control field follows straightforwardly from the equations of motion,
\begin{equation}
Y(s)=\sqrt{u+ \frac{\Delta^2}{4}\sin^2 2\Theta(s)}+ \frac{\Delta}{2}\sin(2\Theta(s)).
\end{equation}

\section*{Equilibrium instantons}

\paragraph*{The equilibrium limit.} Within the optimal control framework, ${\bm \sigma}$ and ${\bm \pi}$ are conjugate variables with Hamiltonian
\begin{equation}\label{eq:ContH}
 \mathcal{H}[{\bm \sigma}, {\bm \pi}]= \frac{1}{2} {\bm \pi}^2 - {\bm \pi} \nabla H({\bm \sigma})-\mu {\bm \sigma}\cdot{\bm \pi}-\frac{z}{2}({\bm \sigma}^2-N).
\end{equation}
As we recall in the SM, this quantity measures how much the dynamical rate \eqref{eq:RescaledPot} decreases if the path was allowed a little more time, i.e., if the time horizon $\hat T$ increases:
\begin{equation}\label{eq:meaningH}
  \frac{\partial\tilde v_{\bm \tau}(q,\hat T)}{\partial\hat T}\Big|_q=-\frac{\mathcal{H}}{N}.
\end{equation}
This quantity then vanishes when the transition is given the optimal amount of time to occur. In that case, one recovers equilibrium processes, i.e., the transitions as they occur spontaneously on their natural, exponentially large timescale. Along the solutions of the FB system \eqref{eq:ContH} reduces to $\mathcal{H}=\frac12\bm \pi\cdot\big(\bm \pi-2\nabla_S H(\bm \sigma)\big)$, where $\nabla_S H$ is the Riemannian gradient on the sphere. This quantity vanishes on two families of paths: free relaxation, $\bm \pi=0$, and its time reversal $\bm\pi=2\nabla_S H$, that is, relaxation in the inverse potential $\dot{\bm \sigma}=\nabla_S H$. Equilibrium instantons are built out of these two solutions. One can check explicitly (see SM) that this is attained in the limit $\Delta\to\infty$: in the reduced variables, $Y(s)=\Delta\sin 2\Theta(s)$ before the saddle and $Y(s)=0$ after it. One can easily check that with our parameterization, the uphill part of this asymptotic solution corresponds exactly to $\bm\pi=2\nabla_S H$, and thus
it is the time reversal of the gradient-flow relaxation towards $\bm \tau$. This asymptotic solution is characterized by detailed balance, which relates the probability of the fluctuating path ${\bm \tau} \to {\bm \sigma}$ and its time-reversed counterpart ${\bm \sigma} \to {\bm \tau}$ by $\mathcal P_{\bm \tau\to \bm \sigma}=\mathcal P^R_{\bm \sigma\to\bm \tau}\,e^{-[H(\bm \sigma)-H(\bm \tau)]/T}$. The rate function \eqref{eq:RescaledPot} for $q\geq 0$ is then recovered from a static large deviation, as the energy cost of equilibrating at zero temperature a configuration ${\bm \sigma}$ under the constraint ${\bm \tau} \cdot {\bm \sigma}=N q$, where ${\bm \tau}$ is the zero temperature equilibrium. That is, from the zero-temperature Franz--Parisi potential \cite{franzRecipesMetastableStates1995a}. For $q<0$ only the
climb to the saddle ($q=0$) is exponentially costly. Explicitly, one then gets \eqref{eq:FranzParisip2}.

\paragraph*{Rates of convergence to equilibrium.} The approach to equilibrium instantons is controlled by how fast the control Hamiltonian vanishes as $\Delta \to \infty$. In the SM, we show that  ${\mathcal{H}}/{N}\sim\Delta^2e^{-2\Delta/n(q)}$ when $\Delta \to \infty$, where $n(q)$ counts the number of times the trajectory passes near stationary points: $n=1$ for $q>0$ (leaving $\bm \tau$), $n=2$ at $q=0$, $n=3$ for $-1<q<0$ (also entering and leaving the saddle), and $n=4$ at $q=-1$. At fixed large $\Delta$, integrating \eqref{eq:meaningH} over the horizon gives 
\begin{equation}\label{eq:Speed}
  \delta\tilde v_{\bm \tau} \simeq
  \begin{cases}
    \delta^{(N)}\,\dfrac{1-q^2}{q^2}\,e^{-2\hat T\delta^{(N)}},                           & 0<q<1,  \\[6pt]
    3\delta^{(N)}\left(\dfrac{q^2}{1-q^2}\right)^{1/3}e^{-\frac{2}{3}\hat T\delta^{(N)}}, & -1<q<0,
  \end{cases}
\end{equation}
with $\delta \tilde v_{\bm \tau}= \tilde v_{\bm \tau}(q,\hat T)-\tilde v_{\bm \tau}(q,\infty)$. The corrections exhibit crossover windows of width $e^{-\hat T\delta^{(N)}/2}$ around $q=0$ and $q=-1$. Eq. \eqref{eq:Speed} shows that at fixed $\delta^{(N)}$, a few relaxation times $1/\delta^{(N)}$ suffice to recover equilibrium instantons. For marginal landscapes, though, this requires $\hat T\gg N^{2/3}$, consistently with the non-commuting limits discussed in the main text. At these large timescales, the exponentially fast convergence is observed at fixed instance, while averaged quantities converge to the equilibrium counterpart only algebraically in $r= \hat T  N^{-2/3}$.

At fixed $s$ (and, for instance, $q=-1$), let $\delta C(s,0) \equiv F_s(\Delta)$ be the difference between the correlation profile and its asymptotic limit (dashed-dotted curve in Fig.~\ref{fig:crossover}). $F_s$ is obtained from the exact solution and decays as $F_s(\Delta)\simeq B(s)\,e^{-a(s)\Delta}$ for $\Delta\to\infty$ (see SM). Since $\Delta^{(N)} = r g_N$, the disorder average reads
\begin{equation}\label{eq:AvC}
\overline{\delta C(s,0)} = \int_0^\infty dg\, P(g)\, F_s(rg) \simeq \frac{b}{r^2}\int_0^\infty dx\, x\, F_s(x).
\end{equation}
For the second equality, we use the fact that for $r\gg1$, the integral is dominated by $g\sim 1/r$, i.e., by instances with an anomalously small gap, where $F_s$ is of order one; \eqref{eq:AvC} follows from the fact that the rescaled-gap distribution $P(g) \stackrel{g \to 0}{\sim} b g$ has a linear tail for small $g$ \cite{perretDensityEigenvaluesSeen2015a, perret2014near}. The averaged profile thus approaches its equilibrium limit only as $1/r^2$, while each instance converges exponentially fast.

\newpage
\appendix
\onecolumngrid
\setcounter{secnumdepth}{3}

\pdfbookmark[0]{Supplemental Material}{SM}
\begin{center}
  {\large\bfseries Supplemental Material}
\end{center}
\vspace{0.5em}

\section{Exact solution of the FB system}\label{app:geodesics}
In this Section of the SM, we derive the exact solution of the Forward-Backward (FB) system of equations for the quadratic random landscapes discussed in the main text. We recall that the system of equations reads
\begin{equation}\label{eq:sigma-pi_SM}
  \begin{split}
    \dot {\bm \sigma}(t)
     & =
    -\mu(t) {\bm \sigma}(t)
    -
    \nabla H( {\bm \sigma}(t))
             +
             {\bm \pi}(t)
    \\
    \dot {{\bm \pi}}(t)
     & =
    \left[
      \mu(t)
      +
      \nabla^2 H( {\bm \sigma}(t))
      \right]{\bm \pi}(t)
    +z(t){\bm \sigma}(t)
  \end{split}
\end{equation}
for $t\in [0, \hat T]$. Here $ H({\bm \sigma})=-\frac12{\bm \sigma}^{\mathsf T}J{\bm \sigma}$ with $J$ a Gaussian matrix, $\mu$ and $z$ are two Lagrange multipliers enforcing the constraints ${\bm \sigma}^2=N$ and ${\bm \sigma}\cdot {\bm \pi}=0$, and the boundary conditions are ${\bm \sigma}(0)= {\bm \tau}$ and ${\bm \pi}(\hat T)=-\tilde{\nu}[{\bm \tau}- N^{-1}({\bm \sigma}(\hat T) \cdot {\bm \tau}){\bm \sigma}(\hat T) ]$. Notice that these are equations of two conjugate variables ${\bm \sigma}$ and ${\bm \pi}$ evolving with conservative dynamics, associated to the Hamiltonian:
\begin{equation}\label{eq:Hamiltonian_SM}
  \mathcal{H}[{\bm \sigma}, {\bm \pi}]= \frac{1}{2} {\bm \pi}^2 - {\bm \pi} \nabla H({\bm \sigma})-\mu {\bm \sigma}\cdot{\bm \pi}-\frac{z}{2}({\bm \sigma}^2-N).
\end{equation}
Along the solutions of \eqref{eq:sigma-pi_SM}, the constraints are satisfied and the Hamiltonian reduces to $\mathcal{H}[{\bm \sigma}, {\bm \pi}] \to   \frac{{\bm \pi}}{2} \left( {\bm \pi}  - 2 \nabla_S H({\bm \sigma}) \right),$ where $\nabla_S H({\bm \sigma})$ indicates the part of the vector $\nabla H({\bm \sigma})$ that is orthogonal to ${\bm \sigma}$ (that is, the Riemannian gradient of $H({\bm \sigma})$ on the sphere). Indeed, only this part survives when projecting onto ${\bm \pi}$ in \eqref{eq:Hamiltonian_SM}, due to the constraint ${\bm \sigma}\cdot{\bm \pi}=0$. Throughout this section, ${\bf v}_\alpha$ are the eigenvectors of $J$ associated to the eigenvalues $\lambda_\alpha$ ($\lambda_1 \leq \cdots \leq \lambda_N$), and we use the notation $\delta_\alpha=\lambda_N -\lambda_{\alpha}$.

\subsection{Mode decomposition of the FB equations}
Owing to the fact that the energy landscape is quadratic, one can decompose the trajectory ${\bm \sigma}(t)$ into the eigenvectors ${\bf v}_\alpha$ of the matrix $J$,  ${\bm \sigma}(t)=\sum_{\alpha=1}^N \sigma_\alpha(t) {\bf v}_\alpha$  with $ \sigma_\alpha(t)={\bm \sigma}(t)\cdot {\bf v}_\alpha$. We adopt the similar decomposition for the control field, ${\bm \pi}(t)=\sum_{\alpha=1}^N \pi_\alpha(t) {\bf v}_\alpha$ with $ \pi_\alpha(t)={\bm \pi}(t)\cdot {\bf v}_\alpha$. The FB system \eqref{eq:sigma-pi_SM}  then corresponds to the following equations for the modes:
\begin{align}
  \dot \sigma_\alpha(t) & = (\lambda_\alpha - \mu(t)) \sigma_\alpha(t) + \pi_\alpha(t),
  \\
  \dot \pi_\alpha(t)    & = \left(\mu(t)-\lambda_\alpha\right) \pi_\alpha(t) + z(t) \sigma_\alpha(t).
\end{align}
Given our choice that ${\bm \sigma}(0)= {\bm \tau}= \sqrt{N} {\bf v}_N$ is one of the two degenerate ground states, then $\sigma_\alpha(0)= \sqrt{N} \delta_{\alpha N}$. The remaining boundary condition reads formally $ \pi_\alpha({\hat T})= -\sqrt{N} \tilde{\nu}(\delta_{\alpha N} - N^{-1}\sigma_N(\hat{T})\sigma_\alpha(\hat{T}) )$. Notice that the equations for different modes are coupled through the Lagrange multipliers, whose solution is
\begin{equation}
  \mu(t) = \frac{1}{N}\sum_{\alpha=1}^N \lambda_\alpha \sigma_\alpha^2(t), \qquad z(t) = -\frac{1}{N} \sum_{\alpha=1}^N \pi_\alpha^2(t).
\end{equation}

This system of coupled equations is closed in the modes, and can be solved by iterating $\sigma_\alpha\to\pi_\alpha\to\sigma_\alpha\to\cdots$ until convergence. From the converged solution, one can compute the observables: 
\begin{align}
  C(t,t')
  =\frac{1}{N}\sum_{\alpha=1}^N\sigma_\alpha(t)\sigma_\alpha(t'), \quad \quad   \epsilon(t)
  =-\frac{1}{2N}\sum_{\alpha=1}^N\lambda_\alpha\sigma_\alpha^2(t).
\end{align}
Examples are given in Fig.~\ref{fig:beta_infty_numerics} in the main text.

\subsection{A family of paths with fixed geometry}
The iteration of the  FB system of equations up to convergence reveals that the optimal escape path belongs to a low dimensional subspace intersected with the sphere: the system chooses a direction to escape from the initial configuration, and this direction is \emph{persistent}: it is not changed along the path. Thus, geometrically the escape paths correspond to arcs on the sphere. Given this observation, one can make an ansatz in which the path moves in the two dimensional space spanned by the initial configuration ${\bm \tau}= \sqrt{N} {\bf v}_N$, and another vector  ${\bf v}^\perp$ of unit norm orthogonal to it, which identifies the direction along which the control field acts at time $t=0$.  Given that the ${\bf v}_\alpha$ form a complete basis, ${\bf v}^\perp=\sum_{\alpha \neq N} v^\perp_{\alpha} {\bf v}_\alpha$. Due to the spherical constraint on ${\bm \sigma}$ and the orthogonality constraint ${\bm \sigma} \cdot {\bm \pi}=0$, the following parametrization holds:
\begin{align}
  {\bm \sigma}(t) & = \sqrt{N}[\cos \theta(t)  {\bf v}_N + \sin \theta(t)  {\bf v}^\perp], \label{eq:FF1}                 \\
  {\bm \pi}(t)    & = \sqrt{N} y(t) \left[-\sin \theta(t) {\bf v}_N + \cos \theta(t) {\bf v}^\perp \right].\label{eq:FF2}
\end{align}
This system is to be solved for $\theta(t), y(t)$ with boundary conditions $\theta(0)=0$ and $y(\hat T)= \tilde{\nu} \sin \theta (\hat T)$, where we are assuming $\sin \theta (\hat T) \neq 0$. Moreover, the direction ${\bf v}^\perp $ is to be determined. Notice that this parametrization is consistent with the dynamical system:  if ${\bm \sigma}(0)$ and ${\bm \pi}(0)$ lie in a subspace spanned by certain eigenvectors of $J$, the dynamics remains in that subspace. \\

We begin by determining the directionality of the path, identified by the vector ${\bf v}^\perp$. To do so, we use the minimum-energy-path (MEP) formulation of Refs.~\cite{eStringMethodStudy2002a,eSimplifiedImprovedString2007}. Plugging the ansatz \eqref{eq:FF1}-\eqref{eq:FF2} in the first of \eqref{eq:sigma-pi_SM}, both $\dot{\bm \sigma}$ and ${\bm \pi}$ are tangent to the path, along $\partial_\theta{\bm \sigma}$. The ansatz is thus consistent only if
\begin{equation}\label{eq:mep_condition_sphere}
  \left(\nabla_S H({\bm \sigma})\right)^\perp=0,
\end{equation}
where the superscript $\perp$ denotes the component normal to the path within the tangent space of the sphere. This is the condition satisfied by an MEP connecting ${\bm \tau}$ and $-{\bm \tau}$.
Define
\begin{equation}
  \bar\lambda=  {\bf v}^\perp\cdot J{\bf v}^\perp,
\end{equation}
and decompose
\begin{equation}
  J{\bf v}^\perp=\bar\lambda\,{\bf v}^\perp+{\bf r},
  \qquad {\bf r}\perp {\bf v}^\perp,\qquad {\bf r}\perp {\bf v}_N.
\end{equation}
The remainder ${\bf r}$ vanishes if and only if ${\bf v}^\perp$ is an eigenvector of $J$. A direct computation gives
\begin{equation}
  -\nabla_S H({\bm \sigma})
  =
  (\bar\lambda-\lambda_N)\sin\theta\cos\theta\,
  \partial_\theta{\bm \sigma}
  +
  \sqrt N\sin\theta\,{\bf r}.
\end{equation}
The first term is tangent to the path, whereas the second term is normal to the
path inside the tangent space of the sphere. Hence
\begin{equation}
  \left(\nabla_S H({\bm \sigma})\right)^\perp
  =
  -\sqrt N\sin\theta\,{\bf r}.
\end{equation}
Therefore the normal component vanishes for all $0<\theta<\pi$ if and only if
${\bf r}=0$, namely if and only if ${\bf v}^\perp$ is an eigenvector of $J$. Thus a great-circle
path through a generic vector ${\bf v}^\perp$ is not a minimum-energy
path: the spherical force has a transverse component that pushes the path out of
the plane spanned by ${\bf v}_N$ and ${\bf v}^\perp$. All possible solutions of the form \eqref{eq:FF1}-\eqref{eq:FF2} are therefore given by  ${\bf v}^\perp={\bf v}_\alpha$, with $\bar\lambda=\lambda_\alpha$. The tangential component then gives $\dot\theta=y-\delta_\alpha\sin\theta\cos\theta$, and projecting the second of \eqref{eq:sigma-pi_SM} onto $\partial_\theta{\bm \sigma}$ gives $\dot y=\delta_\alpha\,y\cos 2\theta$. Explicitly:
\begin{equation}
  \begin{split}
    {\bm \sigma}(t) & = \sqrt{N}[\cos \theta(t)  {\bf v}_N + \sin \theta(t)  {\bf v}_\alpha],                  \\
    {\bm \pi}(t)    & = \sqrt{N} y(t) \left[-\sin \theta(t) {\bf v}_N + \cos \theta(t) {\bf v}_\alpha \right],
  \end{split}
\end{equation}
where the functions $\theta(t), y(t)$ solve the equations (with $\delta_\alpha= \lambda_N -\lambda_\alpha$):
\begin{equation}\label{eq:S1}
  \begin{split}
    \dot \theta(t)= y(t)- \delta_\alpha \sin\theta(t) \cos \theta(t) \quad & \quad \theta(0)=0,                                 \\
    \dot y(t) =  \delta_\alpha  y(t)  \cos [2\theta(t)] \quad              & \quad y(\hat T)= \tilde{\nu} \sin \theta (\hat T).
  \end{split}
\end{equation}
Notice that the Hamiltonian \eqref{eq:Hamiltonian_SM}, evaluated along these paths, reads:
\begin{equation}\label{eq:HamiltonianPaths_SM}
  \mathcal{H}= N \left( \frac{y^2}{2}  -  \delta_\alpha\, y\,  \sin \theta \cos \theta \right)= \frac{N}{2} \left(y^2  -  \delta_\alpha\, y\, \sin (2\theta) \right).
\end{equation}

\begin{figure}[t]
  \centering
  \includegraphics[width=0.48\columnwidth]{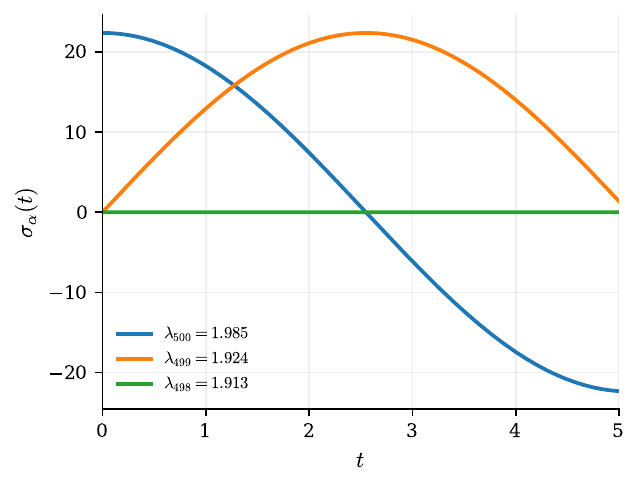}
  \hfill
  \includegraphics[width=0.48\columnwidth]{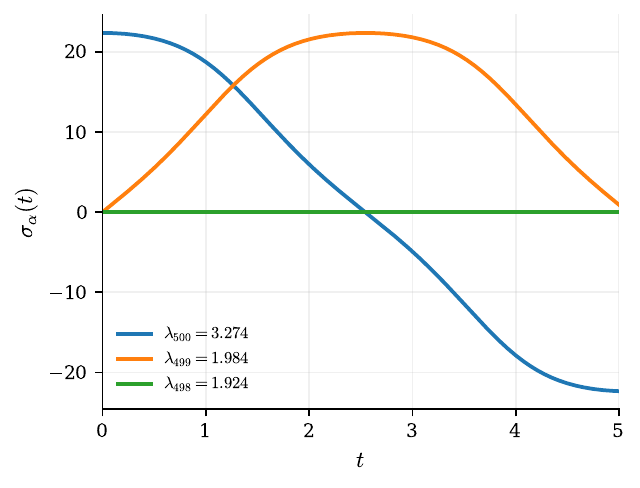}
  \caption{Decomposition of the optimal trajectory into the modes
    $\sigma_\alpha(t)={\bm\sigma}(t)\cdot{\bf v}_\alpha$ associated
    with the three largest eigenvalues, for $J_0=0$ (\emph{Left}) and
    $J_0=3$ (\emph{Right}). Numerical parameters are $N=500$,
    $\hat T=5$, $\tilde\nu=10$ and $N_{\hat T}=500$ time steps. }
  \label{fig:eigenmodes}
\end{figure}

\subsection{The optimal path}
There are $N-1$ possible solutions of the above equations, labeled by $\alpha=1, \cdots, N-1$. The numerical iteration of the FB system of equations selects the solution with $\alpha=N-1$. This is shown in Fig.~\ref{fig:eigenmodes}, which clearly illustrates that  the contribution of the third dominant mode vanishes throughout the interval, and only the first two dominant modes matter. This solution, that corresponds to the lowest energy barrier, is, within the family derived above, the one minimizing the cost functional \eqref{eq:Cost} in the main text. Indeed, for arbitrary $\alpha$, let $\delta\equiv\delta_\alpha$. The cost functional evaluated along the various solutions takes the form
\begin{equation}
  \mathcal{C}_{\hat T, \tilde \nu}[{\bm \pi}; {\bm \tau} ]= \frac{1}{2 N} \int_{0}^{\hat T} dt || {\bm \pi}(t)||^2 +  \frac{\tilde{\nu}}{N}{\bm \sigma}^{\bm \pi}(\hat T)\cdot{\bm \tau} \longrightarrow    \tilde{\mathcal{C}}_{\hat T, \tilde \nu, \delta}[\theta]= \frac{1}{2} \int_{0}^{\hat T} dt    \left(\dot \theta+\delta \sin\theta\cos\theta\right)^2 +  \tilde{\nu}\cos \theta(\hat T).
\end{equation}
Denote by $\theta_\delta$ the minimizing path for a fixed value of $\delta$, and by
\begin{equation}
  \tilde{\mathcal{C}}^*(\delta)=\tilde{\mathcal{C}}_{\hat T, \tilde \nu, \delta}[\theta_\delta]
\end{equation}
the corresponding optimized value. By the envelope theorem, the implicit dependence of $\theta_\delta$ on $\delta$ does not contribute when differentiating $\tilde{\mathcal{C}}^*(\delta)$, because $\theta_\delta$ is stationary. Hence
\begin{align}
  \frac{\dd \tilde{\mathcal{C}}^*}{\dd \delta}
   & =
  \left.
  \frac{\partial \tilde{\mathcal{C}}_{\hat T, \tilde \nu, \delta}[\theta]}{\partial \delta}
  \right|_{\theta=\theta_\delta}
  =
  \int_0^{\hat T}\dd{t}\,
  \sin\theta_\delta(t)\cos\theta_\delta(t)\nonumber
                                          [
                                            \dot\theta_\delta(t)
                                            +
                                            \delta\sin\theta_\delta(t)\cos\theta_\delta(t)
                                          ].
\end{align}
Using that $ \sin\theta_\delta\cos\theta_\delta\,\dot\theta_\delta
  =
  \frac12\frac{\dd}{\dd t}\sin^2\theta_\delta$ and that $\theta_\delta(0)=0$,
we obtain
\begin{align}
  \frac{\dd \tilde{\mathcal{C}}^*}{\dd \delta}
  =
  \left[\frac12\sin^2\theta_\delta(\hat T)\right.\nonumber
    \left.{}+
    \delta
    \int_0^{\hat T}\dd{t}\,
    \sin^2\theta_\delta(t)\cos^2\theta_\delta(t)
    \right]
  \geq 0.
\end{align}
Thus the optimized cost is an increasing function of $\delta$. Among the family of solutions found above, the optimal one is therefore the one with the smallest possible value of $\delta$, namely that with $\alpha=N-1$ and
\begin{equation}
  \delta \to \delta_{N-1}= \lambda_N -\lambda_{N-1}.
\end{equation}
The quantity $ \delta_{N-1}$ is the spectral gap of the matrix $J$; in the main text, we denote it with $\delta^{(N)}$, where the superscript refers to the dimensionality of the matrix $J$.

\subsection{Scaling functions and explicit solution}
\begin{figure}[t]
  \centering
  \includegraphics[width=0.6\textwidth]{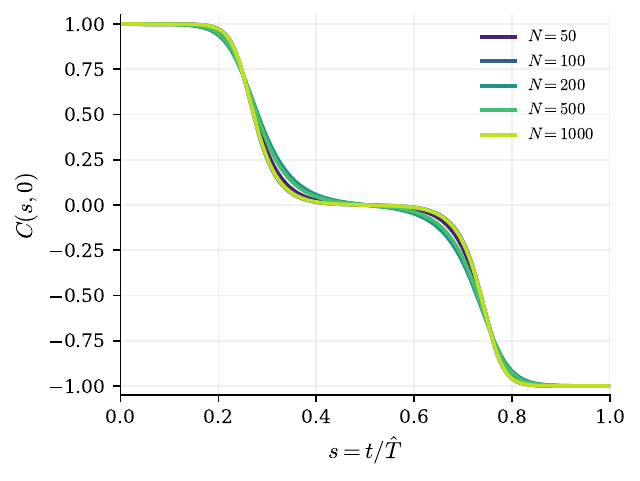}
  \caption{Finite-size scaling of the correlation at fixed $r=10$ and
    $\kappa=100$, for $N=50,100,200,500,1000$.}
  \label{fig:finite_size_collapse}
\end{figure}

The explicit solution given in the main text is obtained upon rescaling of the system \eqref{eq:S1}.  Setting $t= s \hat T$ and defining $\Theta(s)= \theta(s \hat {T})$ and $Y(s)=\hat T y(s \hat {T})$,  as well as $\Delta\equiv  \Delta^{(N)} = \hat{T} \delta^{(N)}$ and $\kappa = \hat{T} \tilde{\nu}$, one gets:
\begin{equation}\label{eq:ParticlePotentialRescaledAppendix}
  \begin{split}
    \frac{d \Theta(s)}{ds}= \hat{T}\frac{d \theta(t)}{dt}\Big|_{t= s\hat{T}}= Y(s)-\frac{\Delta}{2}\sin(2\Theta(s)), \quad & \quad \Theta(0)=0                  \\
    \frac{d Y(s)}{ds}=\hat{T}^2\frac{d y(t)}{dt}\Big|_{t= s\hat{T}}=\Delta\, Y(s)\cos(2\Theta(s)) \quad                    & \quad  Y(1)= \kappa\sin \Theta(1),
  \end{split}
\end{equation}
which are the equations reported in the main text. These equations describe the motion of an effective particle with Hamiltonian 
\begin{equation}\label{eq:HamControlApp}
  \mathcal{H}=\frac{N}{2}(Y^2- \Delta \, Y \, \sin (2 \Theta))= \frac{N}{2} \left(  \dot \Theta^2 - \frac{\Delta^2}{4} \sin^2 (2 \Theta)\right),
\end{equation}
that is the rescaled version of \eqref{eq:HamiltonianPaths_SM}. The quantity $2 \mathcal{H}/N $ is conserved along the trajectory. Thus, we can write (choosing the increasing branch)
\begin{equation}  \label{eq:theta-time-integral}
  \dot \Theta(s)= \sqrt{\frac{2 \mathcal{H}}{N} + \frac{\Delta^2}{4}\sin^2 2\Theta(s)}, \quad \Longrightarrow \quad   s = \int_0^{\Theta(s)} \frac{\dd{\phi}}{\sqrt{\frac{2 \mathcal{H}}{N} + \frac{\Delta^2}{4}\sin^2 2\phi}}.
\end{equation}
Once $\Theta(s)$ is known, one gets
\begin{equation}  \label{eq:Y-time-integral}
  Y(s)=   \dot \Theta(s)+ \frac{\Delta}{2}\sin(2\Theta(s))=  \sqrt{\frac{2 \mathcal{H}}{N} + \frac{\Delta^2}{4}\sin^2 2\Theta(s)}+ \frac{\Delta}{2}\sin(2\Theta(s)).
\end{equation}
Therefore, the explicit solutions for $\Theta(s), Y(s)$ are parametrized by the rescaled energy  $u=2\mathcal{H}/N$, the value of which is fixed by the boundary conditions at the endpoint $s=1$ of the time interval, $\Theta_f:= \Theta(1)$, as:
\begin{equation}\label{eq:HamiltonianDensity}
  u=\frac{2\mathcal{H}}{N}=  \kappa \sin \Theta_f \left( \kappa \sin \Theta_f - \Delta \sin(2 \Theta_f) \right).
\end{equation}
Defining the functions:
\begin{equation}
  \begin{split}
    u_{\Delta, \kappa}(\theta) = \kappa \sin \theta \left( \kappa \sin \theta - \Delta \sin(2 \theta) \right), \quad \quad
    I_{\Delta}(\theta; u)=\int_0^{\theta} \frac{d\phi}{\sqrt{u+ \frac{\Delta^2}{4} \sin^2 2 \phi}}
  \end{split}
\end{equation}
then we see that $\Theta_f$ solves the $\kappa, \Delta$-dependent self-consistent equation:
\begin{equation}
  I_{\Delta}(\Theta_f; u_{\Delta, \kappa}(\Theta_f))=1.
\end{equation}
Then the solutions for $\Theta(s), Y(s)$ are compactly written as
\begin{equation}
  I_{\Delta}(\Theta(s); u_{\Delta, \kappa}(\Theta_f))=s, \quad \quad Y(s)=\sqrt{u_{\Delta, \kappa}(\Theta_f)+ \frac{\Delta^2}{4}\sin^2 2\Theta(s)}+ \frac{\Delta}{2}\sin(2\Theta(s)).
\end{equation}
The solution depends on the disorder only through $\Delta^{(N)}=\hat T\delta^{(N)}$. At fixed $r=\hat T/N^{2/3}$, one has $\Delta^{(N)}=r g_N$ with $g_N=N^{2/3}\delta^{(N)}=O(1)$, so that the profiles at fixed $r$ and $\kappa$ converge to a limiting curve as $N$ increases. This is shown in Fig.~\ref{fig:finite_size_collapse}.

\vspace{1 cm}
\section{Convergence to the equilibrium instantons}
In this part of the SM, we aim at showing that when the limit of infinite time horizon $\hat{T}$ is taken appropriately (meaning, $\Delta \to \infty$), the exact solution of the control problem reproduces the equilibrium instantons. As recalled in the End Matter, the equilibrium instantons are characterized by the fact that the control Hamiltonian vanishes. The Hamiltonian measures the amount the rate function  \eqref{eq:RescaledPot} decreases if the time horizon $\hat T$ increases, see Eq.~\eqref{eq:meaningH} in the End Matter. To see this explicitly, consider the HJB equation \eqref{eq:hjb_terminal} for $T \to 0$, setting $W(t,{\bf x};{\bm \tau})= 2 T V(t,{\bf x};{\bm \tau})$. The HJB equation then reduces to a Hamilton--Jacobi equation
\begin{equation}
  \partial_t W(t,{\bf x};{\bm \tau})=\mathcal{H}[{\bf x},-\nabla_S W(t,{\bf x};{\bm \tau})]
\end{equation}
where $\mathcal{H}$ is \eqref{eq:Hamiltonian_SM} and ${\bm \pi}=-\nabla_S W$ by \eqref{eq:DEPHJB}; the characteristics of this equation are the FB equations. Notice that at $t=0$, $W(0,{\bm \tau};{\bm \tau})=N\,C^*_{\hat T,\tilde\nu}$. Since the controlled dynamics is autonomous, $W$ depends on $t$ and $\hat T$ only through $\hat T-t$, so that $\partial_{\hat T}W=-\partial_tW$. Using the fact that  $\mathcal{H}$  is conserved along the optimal path and exploiting the inverse of the Legendre relation \eqref{eq:LT}, we recover \eqref{eq:meaningH}. The control Hamiltonian converging to zero implies that the time associated to the transition is optimal: this occurs in the limit $\hat T \to \infty$.

In Sec.~\ref{sec:Speed} we show that the control Hamiltonian evaluated along the optimal solutions indeed vanishes exponentially fast in $\Delta$ when $\Delta \to \infty$. We then argue in Sec.~\ref{app:gradient-descent} that solutions with vanishing control Hamiltonian are precisely those that can be built out of time reversal symmetry, and correspond to equilibrium instantons (thus, the convergence to equilibrium instantons is exponentially fast). We show that for these instantons, the dynamical large deviation rate converges to the static Franz-Parisi potential in Sec.~\ref{app:energy}.

\subsection{Speed of convergence to the equilibrium instantons}\label{sec:Speed}
We show here that the control Hamiltonian density $u= 2 \mathcal{H}/N$ evaluated at the solutions of the FB system of equations vanishes exponentially fast in $\Delta$, with coefficients of the exponential decay that depend on $q=\cos \Theta_f$. The speed of convergence is encoded in the endpoint condition $ I_{\Delta}(\Theta_f; u_{\Delta, \kappa}(\Theta_f))=1$, which explicitly reads:
\begin{equation}\label{eq:TermCond}
  1=\int_0^{\Theta_f} \frac{d\phi}{\sqrt{u+ \frac{\Delta^2}{4} \sin^2 2 \phi}}=\frac{1}{\Delta}\int_0^{2\Theta_f} \frac{dx}{\sqrt{\frac{4 u}{\Delta^2}+  \sin^2 x}}
\end{equation}
If  $\frac{4 u}{\Delta^2}$ is small (to be checked self-consistently once we find $u$), this integral is dominated by the region $\sin x \approx 0$, where the denominator is small. This corresponds to $x\sim 0$ (close to the initial condition), $x=\pi$ (close to the saddle) and $x=2 \pi$ (close to the other minimum). Depending on $\Theta_f$, one picks up several of these contributions. Let us distinguish the various cases:

\begin{itemize}
  \item[(1)] \underline{$q \in (0, 1), 0<\Theta_f < \pi/2$.} In this regime, $q$ remains bounded away from zero as $\Delta \to \infty$.  Then the integral \eqref{eq:TermCond} picks up only the contribution from $x \sim 0$. Near this point $\sin^2x\simeq x^2$. We therefore add and subtract the corresponding singular term:
        \begin{equation}
          1=\frac{1}{\Delta}\int_0^{2\Theta_f} \frac{dx}{\sqrt{\frac{4 u}{\Delta^2}+  \sin^2 x}}= \frac{1}{\Delta} \left[ \mathcal{I}_0(2\Theta_f)+\mathcal{R}(2\Theta_f) \right]
        \end{equation}
        where
        \begin{align}\label{Sing1}
          \mathcal{I}_0(2\Theta_f)=\int_0^{2\Theta_f}\frac{dx}{\sqrt{\frac{4 u}{\Delta^2}+x^2}}= \text{arcsinh}\frac{\Theta_f \Delta}{\sqrt{u}} \sim \log \frac{2\Theta_f \Delta}{\sqrt{u}}
        \end{align}
        and the expansion is made assuming $u/\Delta^2 \ll 1$. The regular contribution is
        \begin{equation}
          \mathcal{R}(2\Theta_f)=\int_0^{2\Theta_f}dx\left[\frac{1}{\sqrt{\frac{4 u}{\Delta^2}+\sin^2x}}-\frac{1}{\sqrt{\frac{4 u}{\Delta^2}+x^2}}\right].
        \end{equation}
        This has a well defined limit for $u/\Delta^2 \to 0$, and thus approximately
        \begin{equation}\label{eq:hor_reg}
          \mathcal{R}(2\Theta_f)\simeq\int_0^{2\Theta_f}dx\left[\frac{1}{\sin x}-\frac{1}{x}\right]
          =\Big[\log\tan\tfrac{x}{2}-\log x\Big]_0^{2\Theta_f}=\log\frac{\tan\Theta_f}{\Theta_f}.
        \end{equation}
        Summation of the two contributions gives
        \begin{equation}\label{eq:Upre}
          1=\frac{1}{\Delta} \log \frac{2\Delta \tan\Theta_f }{ \sqrt{u}} \quad  \Longrightarrow \quad u=4 \Delta^2 e^{-2 \Delta} \tan^2 \Theta_f= 4 \Delta^2 e^{-2 \Delta} \frac{1-q^2}{q^2}.
        \end{equation}
        This shows that the control Hamiltonian density vanishes exponentially fast as $\Delta \to \infty$, and thus the hypothesis $ u/\Delta^2 \ll 1$ is consistent. This expression is true provided that $\cos \Theta_f$ remains bounded away from zero, i.e., that $q>0$. The case $q=0$ has to be considered separately, see below.

  \item[(2)] \underline{$q \in (-1,0), \pi/2<\Theta_f <\pi$}. In this regime, $q$ remains negative and strictly larger than $-1$ as $\Delta \to \infty$. In this case, the integral picks contributions from $x \sim 0$ and also $x\sim \pi$. Near these points $\sin^2x\simeq x^2$ and  $\sin^2x\simeq (\pi- x)^2$, respectively. We therefore add and subtract the corresponding singular terms, two in this case:
        \begin{equation}
          1=\frac{1}{\Delta}\int_0^{2\Theta_f} \frac{dx}{\sqrt{\frac{4 u}{\Delta^2}+  \sin^2 x}}= \frac{1}{\Delta} \left[ \mathcal{I}_0(2\Theta_f)+\mathcal{I}_\pi(2\Theta_f)+\mathcal{R}(2\Theta_f) \right]
        \end{equation}
        where $\mathcal{I}_0(2\Theta_f)$ is given in \eqref{Sing1}, while
        \begin{align}\label{IP}
          \mathcal{I}_\pi(2\Theta_f) & =\int_0^{2\Theta_f}\frac{dx}{\sqrt{\frac{4 u}{\Delta^2}+(\pi-x)^2}}=\operatorname{arsinh}\frac{\pi \Delta}{2 \sqrt{u}}-\operatorname{arsinh}\frac{(\pi-2\Theta_f)\Delta}{2 \sqrt u} \sim \log\frac{\pi \Delta^2 (2\Theta_f-\pi)}{ u}
        \end{align}
        where we used again an asymptotic expansion for $ u/\Delta^2 \ll 1$, assuming that $\pi-2\Theta_f$ stays of $O(1)$. The regular part is (setting again $u/\Delta^2 \to 0$):
        \begin{equation}
          \mathcal{R}(2\Theta_f)\approx\int_0^{2\Theta_f}dx\left[\frac{1}{|\sin x|}-\frac{1}{x}-\frac{1}{|x-\pi|}\right].
        \end{equation}
        It is convenient to split this into $[0, \pi]$ and $(\pi, 2\Theta_f]$. The first part gives:
        \begin{equation}\label{eq:PrimP}
          \int_0^{\pi}dx\left[\frac{1}{\sin x}-\frac{1}{x}-\frac{1}{\pi-x}\right]=\Big[\log\tan\tfrac{x}{2}-\log x+\log(\pi-x)\Big]_0^{\pi}=\log\frac{2}{\pi}-\log\frac{\pi}{2}=2\log\frac{2}{\pi},
        \end{equation}
        while the second part gives
        \begin{equation}\label{eq:SP}
          \int_0^{2\Theta_f-\pi}dy\left[\frac{1}{\sin y}-\frac{1}{y}-\frac{1}{\pi+y}\right]
          =\log\frac{2\tan((2\Theta_f-\pi)/2)}{2\Theta_f-\pi}-\log\frac{2\Theta_f}{\pi}.
        \end{equation}
        Combining everything, we find
        \begin{equation}
          \int_0^{2\Theta_f} \frac{dx}{\sqrt{\frac{4 u}{\Delta^2}+  \sin^2 x}}\approx 3\log\frac{2 \Delta}{\sqrt{u}}+\log\tan\frac{2 \Theta_f-\pi}{2}=  3\log\frac{2 \Delta}{\sqrt{u}}+\log|\cot\Theta_f|.
        \end{equation}
        Which leads to
        \begin{equation}\label{eq:Upost}
          1=\frac{1}{\Delta} \left( 3\log\frac{2 \Delta}{\sqrt{u}}+\log|\cot\Theta_f|\right) \quad  \Longrightarrow \quad u= 4 \Delta^2 e^{-\frac{2 \Delta}{3}}(|\cot\Theta_f|)^{\frac{2}{3}}=4 \Delta^2 e^{-\frac{2 \Delta}{3}}\left(\frac{q^2}{1-q^2}\right)^{1/3}.
        \end{equation}
        Again, one has an exponential decay in $\Delta$, but with a different coefficient. This calculation assumes that $q\neq -1$: the case $q=-1$ has to be treated separately, see below.

  \item[(3)] \underline{$q=0, \Theta_f=\pi/2$.} We notice that adapting the calculation of point (2) one can read the case $\Theta_f=\pi/2$. Indeed, in this case the second term in \eqref{IP} does not contribute, nor does \eqref{eq:SP}. Then
        \begin{equation}
          1=\frac{2}{\Delta} \log \frac{2 \Delta}{\sqrt u} \quad \Longrightarrow \quad  u=4 \Delta^2 e^{-\Delta}.
        \end{equation}

  \item[(4)] \underline{$q=-1, \Theta_f=\pi$.} In this case,  the integral also picks up the contribution from the region $x=2\pi$, where $\sin^2x\simeq (2\pi-x)^2$. One has therefore to subtract three singular terms:
        \begin{equation}
          1=\frac{1}{\Delta}\int_0^{2\Theta_f}\frac{dx}{\sqrt{\frac{4u}{\Delta^2}+\sin^2x}}
          =\frac{1}{\Delta}\big[\mathcal{I}_0(2\Theta_f)+\mathcal{I}_\pi(2\Theta_f)+\mathcal{I}_{2\pi}(2\Theta_f)+\mathcal{R}(2\Theta_f)\big].
        \end{equation}
        The first two terms are as in \eqref{Sing1} and \eqref{IP}, and for $\Theta_f=\pi$ they give
        \begin{equation}
          \mathcal{I}_0(2\pi)\sim\log\frac{2\pi\Delta}{\sqrt{u}},\qquad
          \mathcal{I}_\pi(2\pi)\sim2\log\frac{\pi\Delta}{\sqrt{u}} .
        \end{equation}
        The singularity at $x=2\pi$ gives
        \begin{equation}
          \mathcal{I}_{2\pi}(2\Theta_f)=\int_0^{2\Theta_f}\frac{dx}{\sqrt{\frac{4u}{\Delta^2}+(2\pi-x)^2}}
          =\operatorname{arsinh}\frac{\pi\Delta}{\sqrt{u}}-\operatorname{arsinh}\frac{\Delta(\pi -\Theta_f)}{\sqrt{u}}
          \simeq\log\frac{2\pi\Delta}{\sqrt{u}}-\operatorname{arsinh}\frac{\Delta(\pi -\Theta_f)}{\sqrt{u}},
        \end{equation}
        where the second term vanishes for $\Theta_f=\pi$.
        The regular part is bounded, and for $u/\Delta^2\to0$  it converges to
        \begin{equation}
          \mathcal{R}(2\pi)=\int_0^{2\pi}dx\left[\frac{1}{|\sin x|}-\frac{1}{x}-\frac{1}{|x-\pi|}-\frac{1}{2\pi-x}\right]=2\log\frac{2}{\pi^2},
        \end{equation}
        where each of the intervals $[0,\pi]$ and $[\pi,2\pi]$ contributes $2\log\frac{2}{\pi}-\log2=\log\frac{2}{\pi^2}$, using \eqref{eq:PrimP} and $\int_0^\pi dx/(2\pi-x)=\int_0^\pi dy/(\pi+y)=\log2$. Summing everything,
        \begin{equation}\label{eq:u_qm1}
          1=\frac{1}{\Delta}\, 4\log\frac{2\Delta}{\sqrt{u}} \quad\Longrightarrow\quad u=4\Delta^2\,e^{-\Delta/2}.
        \end{equation}

\end{itemize}

We see therefore that the convergence is always exponentially fast in $\Delta$, $u \sim \Delta^2 e^{-\frac{2 \Delta}{n(q)}}$. However, the speed of convergence depends on $q$ through the integer factor $n(q)$. The latter counts the number of times the trajectory passes through a stationary point of the landscape, that is the region where the trajectory spends most of its time: for $q>0$ one has $n(q)=1$ since the trajectory is leaving $\bm {\tau}$, for $q=0$ one has $n(q)=2$ to account for the fact that the trajectory enters the saddle, for $-1<q<0$  one has $n(q)=3$ since the trajectory also leaves the saddle, and finally  for $q=-1$ one has $n(q)=4$ since the trajectory enters the second minimum. \\

These changes in the speed of convergence occur in crossover windows, in which $q$ is exponentially (in $\Delta$) close to $q=0$ and $q=-1$. We now study these crossovers in more detail.

\begin{itemize}
  \item[(a)] \underline{$q=0$ crossover.} Consider $\epsilon_0=\pi/2- \Theta_f$ small, of either sign. The second term resulting from the integral \eqref{IP} suggests to introduce the scaling variable $c_0=\frac{\Delta \epsilon_0}{\sqrt{u}}$. When this quantity is of $O(1)$, the second term resulting from  \eqref{IP} can not be expanded asymptotically, but has to be kept as it is. Following the same calculation steps as above, one then finds
        \begin{equation}\label{eq:u_window}
          1=\frac{1}{\Delta}\left(2\log\frac{2\Delta}{\sqrt{u}}-\operatorname{arsinh}\frac{\Delta \epsilon_0}{\sqrt{u}}\right),
        \end{equation}
        which defines $u$ as a function of $\Delta$ and $\epsilon_0$. The $q=0$ scaling is recovered for $c_0 \to 0$, while the $q\neq 0$ regimes are obtained for $c_0 \to \pm \infty$.

  \item[(b)] \underline{$q=-1$ crossover.} In this case one considers $\epsilon_1=\pi- \Theta_f$ small, positive. The scaling variable is again $c_1=\frac{\Delta \epsilon_1}{\sqrt{u}}$. With the same exact reasoning as above, one gets the relation:
        \begin{equation}\label{eq:u_window_qm1}
          1=\frac{1}{\Delta}\left(4\log\frac{2\Delta}{\sqrt{u}}-\operatorname{arsinh}\frac{\Delta \epsilon_1}{\sqrt{u}}\right).
        \end{equation}
        Notice that the same calculation gives the time $s_{\pi/2}$ at which the trajectory crosses the saddle. By symmetry of the integrand in \eqref{eq:TermCond} around $\pi/2$ and $\pi$, the time spent between the saddle and the endpoint equals the time from the saddle to $\pi$ minus the time from $\Theta_f$ to $\pi$, so that
        \begin{equation}\label{eq:Taddle}
          2s_{\pi/2}-1=\int_{\Theta_f}^{\pi}\frac{d\phi}{\sqrt{u+\frac{\Delta^2}{4}\sin^22\phi}}\simeq\frac{1}{\Delta}\operatorname{arsinh}c_1,
          \qquad
          s_{\pi/2}=\frac12+\frac{\operatorname{arsinh}c_1}{2\Delta}.
        \end{equation}
        Within the crossover window $c_1=O(1)$, so that $s_{\pi/2}\to1/2$: the instanton becomes symmetric around the saddle-crossing time, and the uphill branch is the time reversal of the downhill one.
\end{itemize}

\subsection{Asymptotic $\Delta \to \infty$ solution, time reversal and shape of the instantons}\label{app:gradient-descent}

In the previous section we proved that the control Hamiltonian density $u$ vanishes exponentially fast when $\Delta \to \infty$, for any value of $q$. This immediately implies, by Eq.~\eqref{eq:Y-time-integral}, that in the limit
\begin{equation}\label{eq:AsyControl}
  Y(s) \sim \begin{cases}
    \Delta \sin[2\Theta(s)],
     & s<s_{ \pi/2}, \\
    0,
     & s>s_{ \pi/2},
  \end{cases} \quad  \Longrightarrow  \quad
  y(t) =
  \begin{cases}
    \delta^{(N)} \sin [2\theta(t)],
     & t<t_{ \pi/2}, \\
    0,
     & t>t_{ \pi/2}
  \end{cases}
\end{equation}
where $t_{ \pi/2}= s_{ \pi/2} \hat T$ is the time at which the saddle ($\theta=\pi/2$) is reached. This holds for any $\Theta(1)$: if $\Theta(1)<\pi/2$ (i.e., $q>0$) the saddle is not reached within the horizon, $s_{\pi/2}>1$, and only the first branch is relevant.

The first part is the fluctuating path, and it is the motion of a particle that evolves in the inverse potential, i.e., following $\dot{\bm \sigma}= \nabla_{S}H({\bm \sigma})=\nabla H({\bm \sigma})+\mu(t) {\bm \sigma}$.  Indeed,
\begin{equation}
  \nabla_{S}H({\bm \sigma})=\nabla H({\bm \sigma})+\mu(t) {\bm \sigma}= - J {\bm \sigma} + \mu(t) {\bm \sigma}.
\end{equation}
In our parametrization this is
\begin{equation}
  \begin{split}
    \nabla_{S}H({\bm \sigma}) & = - \sqrt{N}[\cos \theta(t) (\lambda_N-\mu(t))  {\bf v}_N + \sin \theta(t) (\lambda_{N-1}-\mu(t) ) {\bf v}_{N-1}]             \\
                              & = - \sqrt{N}[\cos \theta(t) \sin^2 \theta \delta^{(N)}  {\bf v}_N - \sin \theta(t) \cos^2 \theta \delta^{(N)}  {\bf v}_{N-1}] \\
                              & =  \sqrt{N} \cos \theta(t) \sin \theta \delta^{(N)}  [-  \sin \theta {\bf v}_N + \cos \theta {\bf v}_{N-1}],
  \end{split}
\end{equation}
which shows that indeed $y(t)= \delta^{(N)} \, \sin 2\theta $ and ${\bm \pi}= 2 \nabla_{S}H({\bm \sigma})$. This is consistent with $u=0$: the control Hamiltonian \eqref{eq:Hamiltonian_SM} vanishes for ${\bm \pi}= 2 \nabla_{S}H({\bm \sigma})$ (when ${\bm \pi} \neq 0$), and therefore the controlled dynamics at $u=0$ is equivalent to the dynamics in the inverted potential. Such motion is the time reversal of gradient descent, i.e., of a particle evolving under $\dot{\bm \sigma}= -\nabla_{S}H({\bm \sigma})=-\nabla H({\bm \sigma})-\mu(t) {\bm \sigma}$. \\

The asymptotic solution \eqref{eq:AsyControl} determines the shape of the optimal path explicitly. On the ascending branch, inserting $Y=\Delta\sin2\Theta$ into $\dot\Theta=Y-\frac{\Delta}{2}\sin2\Theta$ gives $\dot\Theta=\frac{\Delta}{2}\sin2\Theta$, i.e.\ $\frac{d}{ds}\tan\Theta=\Delta\tan\Theta$. Denoting by $s_{\pi/4}$ the time at which $\Theta(s_{\pi/4})=\pi/4$,
\begin{equation}
  \tan\Theta(s)=e^{\Delta(s-s_{\pi/4})},\qquad
  Y(s)=\Delta\sin2\Theta=\frac{2\Delta\tan\Theta}{1+\tan^2\Theta}=\frac{\Delta}{\cosh[\Delta(s-s_{\pi/4})]}=2\dot\Theta(s).
  \label{eq:kink_up}
\end{equation}
The trajectory is thus a kink of width $1/\Delta$, steepest at $\Theta=\pi/4$, and the control is a pulse localized at the same time, carrying the whole cost of the ascent. On the descending branch $Y=0$, and the path follows free gradient descent, $\dot\Theta=-\frac{\Delta}{2}\sin2\Theta$. Setting $\Theta=\frac{\pi}{2}+\varphi$, this becomes $\dot\varphi=\frac{\Delta}{2}\sin2\varphi$, so that
\begin{equation}
  \tan\!\Big(\Theta(s)-\frac{\pi}{2}\Big)=e^{\Delta(s-s_{3\pi/4})},\qquad Y(s)=0,
  \label{eq:kink_down}
\end{equation}
with $s_{3 \pi/4}$ the time at which $\Theta=3\pi/4$. The position of the kinks follows from the horizon constraint: $s_{\pi/4}$ is the time needed to reach $\Theta=\pi/4$, i.e.\ one quarter period, $s_{\pi/4}=\frac{1}{\Delta}\log\frac{2\Delta}{\sqrt{u}}$, and by symmetry $s_{\pi/2}=2s_{\pi/4}$, $s_{3\pi/4}=3s_{\pi/4}$. Using the values of $u$ derived in Sec.~\ref{sec:Speed},
\begin{equation}\label{eq:CrossingTimes}
  s_{\pi/4}=
  \begin{cases}
    1-\frac{1}{\Delta}\log\tan\Theta(1),                   & q>0,    \\
    \frac12,                                               & q=0,    \\
    \frac13+\frac{1}{3\Delta}\log\frac{\sqrt{1-q^2}}{|q|}, & -1<q<0, \\
    \frac14,                                               & q\to-1,
  \end{cases}
  \qquad
  s_{3\pi/4}=
  \begin{cases}
    1-\frac{1}{\Delta}\log\frac{|q|}{\sqrt{1-q^2}}, & -1<q<0,  \\[2pt]
    \frac34,                                        & q\to-1 .
  \end{cases}
\end{equation}
These formulas hold for any $\Theta(1)$, even when $s_{\pi/4}$, $s_{\pi/2}$ or $s_{3\pi/4}$ exceed $1$: the corresponding portion of the trajectory is then simply not reached within the horizon. For instance, for $\Theta(1)<\pi/4$ one has $s_{\pi/4}>1$: the ascending kink is truncated before its steepest point, and the control grows monotonically up to the final time, $Y(1)=\Delta\sin2\Theta(1)$. Similarly, for $\pi/2<\Theta(1)<3\pi/4$ one has $s_{3\pi/4}>1$, and the descending kink is truncated. Eqs.~\eqref{eq:kink_up}--\eqref{eq:kink_down} hold up to corrections of order $\sqrt{u}/\Delta$, which are exponentially small in $\Delta$ and are relevant only in the scaling windows close to the stationary points.

In terms of the overlap and of the energy density, these formulas imply
\begin{equation}\label{eq:CorrLargeDelta}
  C(s,0)=\frac{1}{\sqrt{1+e^{2\Delta(s-s_{\pi/4})}}}-\frac{1}{\sqrt{1+e^{-2\Delta(s-s_{3\pi/4})}}}
\end{equation}
and 
\begin{equation}
  \epsilon(s)-\epsilon(\bm \tau)=\frac{\delta}{2}\left[\frac{1}{1+e^{-2\Delta(s-s_{\pi/4})}}-\frac{1}{1+e^{-2\Delta(s-s_{3\pi/4})}}\right],
\end{equation}
so that the energy grows to the barrier value $\delta/2$ through a logistic step centred at $s_{\pi/4}$, and decreases through a second one centred at $s_{3\pi/4}$. Outside the kinks the trajectory lies on plateaux at the stationary points (initial minimum, saddle, opposite minimum), where the landscape force vanishes and the system lingers at no cost.

\paragraph{Approach of the correlation profile to its asymptotic form.}
Eq.~\eqref{eq:CorrLargeDelta} holds up to corrections that are relevant only when the angle is within $O(\sqrt{u}/\Delta)$ of a stationary point. Therefore \eqref{eq:CorrLargeDelta} can be used to compute the deviation $\delta C(s,0)=C(s,0)-C_\infty(s)$ from the asymptotic plateau profile $C_\infty(s)$ for all $s$, except at the times at which the path crosses a stationary point. 
Expanding \eqref{eq:CorrLargeDelta} on each plateau, we find $\delta C(s,0)\simeq B_q(s)\,e^{-a_q(s)\Delta}$ with coefficients that depend on the chosen $q$. For $q=-1$, in particular, it holds $C_\infty(s)=1,\,0,\,-1$ for $s<1/4$, $1/4<s<3/4$, $s>3/4$, and:
\begin{equation}
a_{q=-1}(s)=\begin{cases} 2\left(\tfrac14-s\right), \\ s-\tfrac14, \\ \tfrac34-s, \\ 2\left(s-\tfrac34\right), \end{cases}
\qquad
B_{q=-1}(s)=\begin{cases} -\tfrac12, & s<\tfrac14,\\ 1, & \tfrac14<s<\tfrac12,\\ -1, & \tfrac12<s<\tfrac34,\\ \tfrac12, & s>\tfrac34. 
\end{cases}
\end{equation}
The exponent $a_{q=-1}(s)$ is twice the distance from the nearest kink on the plateaux at the minima, and the distance from the nearest kink on the plateau at the saddle. The formulas fail at the saddle-crossing time $s=1/2$, and in a boundary layer of width $1/\Delta$ near $s=1$.

\vspace{1 cm}

\paragraph*{Gradient descent from the saddle: finite $\Delta$ as a kick.}
The descending branch \eqref{eq:kink_down} can be read as a relaxation from the saddle. To see this, we study the gradient-descent dynamics initialized near the saddle ${\bf s}_{N-1}=\sqrt N\,{\bf v}_{N-1}$. An initial condition placed exactly at the saddle would remain there, so we introduce a small perturbation toward the opposite minimum:
\begin{equation}
  {\bm \sigma}(0)=\cos\varepsilon\,{\bf s}_{N-1}-\sin\varepsilon\,{\bf s}_N,
  \qquad \varepsilon\ll1.
\end{equation}
Under gradient descent, the system remains in the plane spanned by ${\bf s}_{N-1}$ and ${\bf s}_N$. Setting $y=0$ and $\theta(0)=\varepsilon$ gives
\begin{align}
  {\bm \sigma}(t') & =\cos\theta(t')\,{\bf s}_{N-1}-\sin\theta(t')\,{\bf s}_N, \\
  \dot\theta       & =\delta\sin\theta\cos\theta,
  \qquad \delta=\lambda_N-\lambda_{N-1},
\end{align}
where $t'$ is the time elapsed since leaving the saddle. This equation has the explicit solution
\begin{equation}\label{eq:GDsaddle}
  \tan \theta(t') = \tan(\varepsilon)\, e^{\delta t'}.
\end{equation}
With $t'=(s-s_{\pi/2})\hat T$ and $\theta=\Theta-\pi/2$, this coincides with the descending branch \eqref{eq:kink_down}, provided that
\begin{equation}\label{eq:kick}
  \tan\varepsilon=e^{-\Delta(s_{3\pi/4}-s_{\pi/2})}=e^{-\Delta s_{\pi/4}}=\frac{\sqrt{u}}{2\Delta}.
\end{equation}
Similarly, the ascending branch \eqref{eq:kink_up} can be written as $\cot\Theta(s)=\tan(\varepsilon)\,e^{\Delta(s_{\pi/2}-s)}$, i.e., it is the time reversal of the same relaxation. Therefore, a finite $\Delta$ (finite $\hat T$) acts as a kick: at $u=0$ the trajectory would take an infinite time to leave the saddle, while for $u>0$ it crosses it with finite velocity $\dot\Theta(s_{\pi/2})=\sqrt u$. Away from the saddle, i.e., for $|\Theta-\pi/2|\gg\sqrt u/\Delta$, the effect of this crossing is equivalent to starting the relaxation at a distance $\varepsilon$ from the saddle. Using the values of $u$ derived in Sec.~\ref{sec:Speed}, the kick is exponentially small in $\Delta$:
\begin{equation}
  \varepsilon\simeq
  \begin{cases}
    e^{-\Delta/3}\left(\frac{|q|}{\sqrt{1-q^2}}\right)^{1/3}, & -1<q<0, \\
    e^{-\Delta/4},                                            & q=-1.
  \end{cases}
\end{equation}

\subsection{Convergence to the static rate: the Franz-Parisi Potential}\label{app:energy}
Plugging the asymptotic solution \eqref{eq:AsyControl} into the definition of $\tilde v_{\bm \tau}(q, \hat T)$, Eq. \eqref{eq:RescaledPot}, immediately gives \eqref{eq:FranzParisip2}. The dynamical large deviation function $\tilde v_{\bm \tau}(q, \hat T)$ therefore converges (for $\hat {T} \to \infty$ and $q\geq 0$) to the function $\tilde v_{\rm FP}(q) =2 T  v_{\mathrm{FP}}(q)$ when $q\geq 0$, where $ v_{\mathrm{FP}}(q)$ is the Franz-Parisi potential associated to our model. The latter is defined as the (free)-energy density cost required to equilibrate a configuration ${\bm \sigma}$ at temperature $T$ and  at fixed overlap $q$ from another reference configuration ${\bm \tau}$ extracted from an equilibrium measure at temperature $T_0$. If $T=T_0$, this is the difference between the free-energy density of the configuration equilibrated under the constraint ${\bm \sigma} \cdot {\bm \tau}=N q$, and that without constraint (that is, the free-energy density of the reference configuration ${\bm \tau}$). When $T \to 0$, this becomes a difference in energy densities.

For the spherical $(p=2)$-spin model that we are considering, this potential is very simple to compute. For $T \to 0$, the reference configuration extracted at equilibrium is one of the two degenerate ground states, i.e., ${\bm \tau}={\bf s}^+_N=\sqrt N\,{\bf v}_N$. Writing ${\bf x}={\bm \sigma}/\sqrt N$, any configuration with
overlap $q={\bf x}\cdot {\bf v}_N$ has the form $ {\bf x}=q {\bf v}_N+\sqrt{1-q^2}\,{\bf y}$ with $  {\bf y}\perp {\bf v}_N$ and $  \|{\bf y}\|=1$.
The energy density is
\begin{equation}
  \frac{H({\bm \sigma})}{N}
  =-\frac{1}{2}
  \left[
    \lambda_Nq^2
    +(1-q^2){\bf y}^{\mathsf T}\, J {\bf y}
    \right].
  \label{eq:energy_fixed_overlap}
\end{equation}
The minimum  of this function is obtained for ${\bf y}=\pm {\bf v}_{N-1}$, so the minimizing configurations are $ {\bf x}^\pm_q=q {\bf v}_N\pm\sqrt{1-q^2}\,{\bf v}_{N-1}$.
Subtracting the ground-state energy density gives the static, zero temperature Franz--Parisi potential $T v_{\mathrm{FP}}(q)=\frac{\delta^{(N)}}{2}(1-q^2)$, which coincides with  $\tilde v_{\rm FP}(q)$ up to a factor of two that corresponds to our definition of $\tilde v_{\bm \tau}(q, \hat T)$.

This convergence is expected, and it is a consequence of the fact that the trajectories in the limit $\hat T \to \infty$ are equilibrium-like, and thus satisfy detailed balance. The latter relates the probabilities of the fluctuating path ${\bm \tau} \to {\bm \sigma}$ and its time-reversed counterpart ${\bm \sigma} \to {\bm \tau}$ by $\mathcal P_{\tau\to\sigma}=\mathcal P^R_{\sigma\to\tau}\,e^{-[H(\sigma)-H(\tau)]/T}$, where $H$ is the system's Hamiltonian. Since the relaxation path ${\bm \sigma} \to {\bm \tau}$ is typical with gradient descent dynamics, the corresponding probability $\mathcal P^R_{\sigma\to\tau} \sim O(1)$ (not exponential in $N, 1/T$) and thus
\begin{equation}
  \tilde v_{\bm \tau}(q, \hat T \to \infty)=- \frac{2T}{N} \log \mathcal P_{\tau\to\sigma} \sim   \frac{2}{N} \left[ H({\bm \sigma})-H({\bm \tau}) \right] = \tilde v_{\rm FP}(q) \quad \quad q\geq 0.
\end{equation}
When $q<0$, the exponential cost of the path is associated to the fluctuation part only, that goes from ${\bm \tau}$ up to the saddle. The corresponding difference in energy density is thus proportional to the spectral gap $\delta^{(\infty)}$, matching with \eqref{eq:FranzParisip2}.

\section{The biased ensemble and the Legendre transform}\label{sec:KappaScaling}
In this section we investigate the behaviour of the biased ensemble, introduced to measure large deviations of the dynamical paths. In Sec. \ref{sec:pinning} we discuss a pinning transition occurring in the ensemble. In  Sec. \ref{sec:BiasScaling} we investigate how the ensemble behaves when $\Delta \to \infty$, determining what is the scaling of the parameter $\kappa$ that biases the trajectory ensemble and allows to reconstruct the dynamical rate function $\tilde v_{\bm \tau}(q, \hat T)$. We notice that physically, the bias acts as a force conjugate to the overlap, see \eqref{eq:LT}. Inversion of the Legendre transform gives $\tilde\nu=-\partial_q\tilde v_{\bm \tau}(q,\hat T)$. When $\Delta \to \infty$ ($\hat T \to \infty$ sufficiently fast with $N$), $\tilde v_{\bm \tau}(q,\hat T)$ reaches the limit \eqref{eq:FranzParisip2}, which in terms of the scaling variables gives asymptotically:
\begin{equation}\label{eq:FranzParisip222}
  \frac{\kappa}{\Delta} \stackrel{\Delta \to \infty}{\longrightarrow}
  \begin{cases}
    2 q, & 0< q< 1,  \\
    0,   & -1< q< 0.
  \end{cases}
\end{equation}
Therefore, the biased ensemble does not allow to recover trajectories with endpoint at $-1<q<0$ in this limit. In fact, the only trajectories that are recovered in this scaling are those reaching endpoint conditions asymptotically close to $q=-1$, as we show below. This is because the limiting function  \eqref{eq:FranzParisip2} is flat in  $-1<q<0$, and it is concave (not convex!) for $0< q< 1$: the Legendre transform thus recovers the convex hull of  \eqref{eq:FranzParisip2}, that consists of the two points $q=\pm1$ only.

\subsection{Pinning transition in the biased ensemble}\label{sec:pinning}
\begin{figure}[t]
  \centering
  \includegraphics[width=.48\columnwidth]{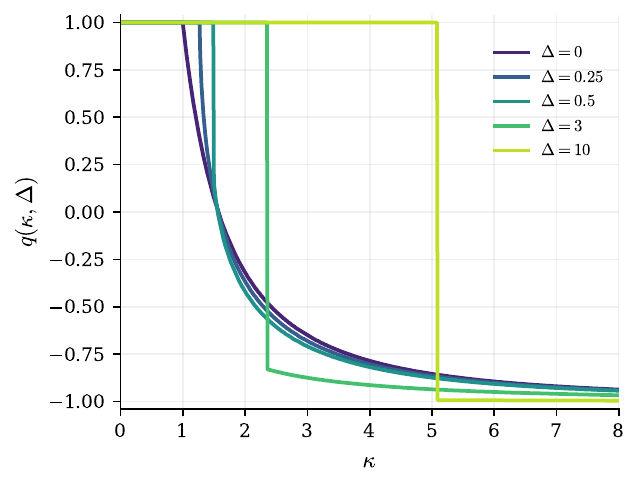}
  \includegraphics[width=.48\columnwidth]{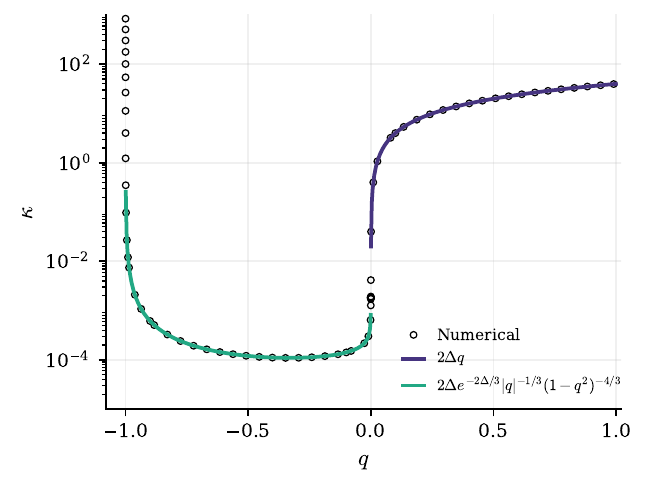}
  \caption{\textit{Left.} Pinning transition of the final overlap
    $q(\kappa,\Delta)$ as the rescaled tilt $\kappa$ is varied, for several
    values of $\Delta$. The pinning transition is continuous at
    small $\Delta$ and becomes discontinuous as $\Delta \to \infty$.
    \textit{Right.} Tilt $\kappa$ needed to reach the final overlap $q$, at $\Delta=20$.
    Points are obtained by solving $I_\Delta(\arccos q; u)=1$ numerically for $u$ and
    setting $\kappa=Y(1)/\sin\Theta(1)$. Lines are the asymptotic predictions of
    Sec.~\ref{sec:BiasScaling}: $\kappa=2\Delta q$ for $0<q<1$, see \eqref{eq:kappaIn1},
    and \eqref{eq:AA} for $-1<q<0$. The crossovers at $q=0$ and $q=-1$ occur in
    windows of width $e^{-\Delta/2}$, not resolved on this scale.}
  \label{fig:kappa_crossover}
\end{figure}

In the large deviation approach, the dynamical rate $\tilde v_{\bm \tau}(q,\hat T)$ is obtained through a Legendre transform of $ \tilde \psi_{\bm \tau}(\tilde \nu, \hat T) $, which governs the asymptotics of the generating function \eqref{eq:GenF}. The latter is an average over an ensemble of trajectories biased by the exponential weight with strength $\tilde \nu$. In the problem we are considering, a too-small $\tilde \nu$  (or, after rescaling, a too-small $\kappa$)  is not sufficient to produce an atypical trajectory. At weak tilt, the optimal trajectory remains pinned at the typical solution $q=1$, i.e., the ensemble does not select atypical paths. This is immediately evident at $\Delta=0$. In this case, the equations of motion yield $Y=\kappa \sin \Theta(1)$ constant, and $\Theta(s)= Y s= \kappa \sin \Theta(1) s$. Plugging $s=1$, one gets the self-consistent equation $\Theta(1)= \kappa \sin \Theta(1)$, which has $\Theta(1)=0$ (corresponding to $q=1$) as a unique solution for all $\kappa\leq \kappa_c=1$. Therefore, the biased ensemble for $\Delta=0$ is pinned to the typical trajectories for bias smaller than $\kappa_c$. For $\Delta>0$, the same phenomenology occurs, with a $\kappa_c(\Delta)$. Figure~\ref{fig:kappa_crossover} shows the final overlap $q$ selected by the tilted action as a function of $\kappa$ for several fixed values of $\Delta$. Increasing $\kappa$ produces a continuous departure from the pinned solution at small $\Delta$, and a steep jump at  $\Delta$ large, from  $q=1$ (the typical trajectory) to $q=-1$ (transition between the two minima). Below, we characterize this behavior and show that $\kappa_c(\Delta)/\Delta \to 1/2$ for $\Delta \to \infty$.

\subsection{Scaling of the bias $\kappa$ and its interpretation}\label{sec:BiasScaling}
The scaling of $\kappa$ is obtained from the boundary condition $Y(1)=\kappa \sin \Theta_f$. We make use of the fact that, for arbitrary $\Delta$ and for  $ u \ll \Delta^2 |\sin (2\Theta_f)|^2$,
\begin{equation}\label{eq:ToExp}
  \begin{split}
    Y(1) & = \sqrt{u+ \frac{\Delta^2}{4}\sin^2 (2\Theta_f)}+ \frac{\Delta}{2}\sin(2\Theta_f) \approx
    \frac{\Delta}{2} \sin(2\Theta_f) + \frac{\Delta}{2} |\sin(2\Theta_f)|+\frac{u}{\Delta |\sin (2\Theta_f)|}
  \end{split}
\end{equation}
Again, we consider different regimes for $q=\cos \Theta_f$.

\begin{itemize}
  \item[(1)]  \underline{$q \in (0, 1), 0 \leq \Theta_f<\pi/2$}. In this case, there is only the fluctuating part of the path. The $\Delta \to \infty$ solution yields $Y(1) = \Delta \sin(2\Theta_f)$, and thus $\kappa=\Delta \sin(2\Theta_f)/ \sin \Theta_f=2 \Delta \cos (\Theta_f)=2 \Delta q$. We can compute corrections to this scaling via the expansion \eqref{eq:ToExp}, which gives
        \begin{equation}\label{eq:kappaIn1}
          \kappa \approx 2 \Delta q +\frac{u}{2 \Delta (1-q^2)q}= 2\Delta q+\frac{2\Delta\,e^{-2\Delta}}{q^3},
        \end{equation}
        where we have used \eqref{eq:Upre}. For $q>0$ remaining bounded away from zero as $\Delta \to \infty$, the second term represents an exponentially small correction, and the bias $\kappa$ grows linearly with $\Delta$. This is easily interpretable: in this regime, when $\hat T \to \infty$ the path climbs by reversed relaxation, and it must be held at $q$ against the drift pulling it back to ${\bm \tau}$.   The bias equals the restoring force of the landscape at that point, i.e. the slope of the static Franz–Parisi potential $\delta^{(N)}(1-q^2)$. This fluctuation thus probes the constrained energy landscape. Notice that this expansion assumes $u \ll \Delta^2 |\sin (2\Theta_f)|^2$. Using again \eqref{eq:Upre}, we see that this corresponds to $q=\cos \Theta_f \gg e^{-\Delta/2}$. Therefore, in the scaling regime $q \sim e^{-\Delta/2}$, in which the endpoint is exponentially close to the saddle, \eqref{eq:kappaIn1} breaks down and the scaling of $\kappa$ has to be analyzed with more care.\\

  \item[(2)] \underline{$q \in (-1,0), \pi/2<\Theta_f<\pi$}. In this case there is the fluctuation path up to the saddle, plus the gradient descent to the right $q$. The $\Delta \to \infty$ solution gives $Y(1)=0$. Since $\sin \Theta_f \neq 0$, then the boundary condition implies $\kappa=0$ always. In fact, the expansion \eqref{eq:ToExp} gives  $Y(1) \approx u/(\Delta |\sin (2\Theta_f)|)$ and
        \begin{equation}\label{eq:AA}
          \kappa \approx \frac{u}{2 \Delta (1-q^2)|q|}= \frac{2 \Delta e^{-\frac{2 \Delta}{3}}}{ (1-q^2)^{4/3}|q|^{1/3}}
        \end{equation}
        where we have used \eqref{eq:Upost}. Therefore, in these regimes the bias $\kappa$ is exponentially small in  $\Delta$. This exponentially small bias is purely a timing cost: it is the price of arriving at that particular $q$ exactly at $\hat T$, and it vanishes as $\hat T \to \infty$. As above, this is derived assuming that $u \ll \Delta^2 |\sin (2\Theta_f)|^2$, that means $|q| (1-q^2) \gg e^{-\Delta/2}$. The scaling  regimes $|q| \sim e^{-\Delta/2}$ and $1+q \sim  e^{-\Delta/2}$ have to be analyzed with more care.\\

  \item[(3)]\underline{Scaling at $q=0$. } As $q$ crosses zero, the tilt $\kappa$ goes from linear to exponentially small in $\Delta$. Exactly at  $q=0, \Theta_f=\pi/2$, plugging these values in  \eqref{eq:ToExp} before performing any expansion leads to $\kappa=\sqrt{u}= 2 \Delta e^{-\frac{\Delta}{2}}$. Exactly at the saddle, $\kappa=\sqrt u$: the only force needed is the one that brings the system to the top in time. The scaling suggests to introduce the scaling variable $\hat \kappa$ with $\kappa= 2 \Delta e^{-\frac{\Delta}{2}} \hat \kappa$. The endpoint condition can no longer be expanded as in \eqref{eq:ToExp}. With $\epsilon_0=\pi/2 -\Theta_f$ and $c_0=\Delta \epsilon_0/\sqrt u$ and using $q=\sin\epsilon_0\simeq\epsilon_0$ and $\sin\Theta_f\simeq 1$, we get instead:
        \begin{equation} \label{eq:kappa_window}
          \kappa\simeq Y(1)=\sqrt{u+\Delta^2\epsilon_0^2}+\Delta\epsilon_0 =\sqrt{u}\,\big(c_0+\sqrt{1+c_0^2}\big)=\sqrt{u}\,e^{\operatorname{arsinh} c_0}
        \end{equation}
        Combining this with \eqref{eq:u_window} gives
        \begin{equation}
          \kappa=2\Delta\,e^{-\Delta/2}\,e^{\frac12\operatorname{arsinh}c_0} \equiv 2\Delta\,e^{-\Delta/2}\,\hat \kappa, \quad \quad u=\frac{4\Delta^2e^{-\Delta}}{\hat\kappa^2}.
        \end{equation}
        Given that $c_0=(\hat \kappa^2-\hat \kappa^{-2})/2$,  we recover the fact that the scaling window is exponentially small:
        \begin{equation}
          q=\epsilon_0=\frac{c_0\sqrt{u}}{\Delta}=2c_0\,e^{-\Delta/2}\,e^{-\frac12\operatorname{arsinh}c_0}= \,e^{-\Delta/2}(\hat\kappa-\hat\kappa^{-3}).
        \end{equation}
        These equations hold for both signs of $\epsilon_0$, and describe the whole window around $q=0$. This scaling matches with the regimes analyzed above. In particular, for $\hat\kappa\to \infty$ one gets $\hat \kappa= q e^{\Delta/2}+ (q e^{\Delta/2})^{-3}$ and one recovers \eqref{eq:kappaIn1},  $\hat \kappa=1$ gives the $q=0$ case and  $\hat\kappa\to 0$ gives $\hat \kappa = (|q| e^{\Delta/2})^{-1/3}$ and thus  \eqref{eq:AA}. \\

  \item[(4)] \underline{Scaling at $q=-1$. } Also  $q=-1, \Theta_f=\pi$ plugged in  \eqref{eq:ToExp} before performing any expansion leads to $Y(1)=\sqrt{u}= 2 \Delta e^{-\frac{\Delta}{4}}$, but $Y(1)= \kappa \sin \Theta_f$ with  $ \sin \Theta_f=0$. Therefore, $\kappa \to \infty$ at this point.  We now set $\epsilon_1=\pi- \Theta_f$, with $ q=-\cos\epsilon_1\simeq-1+\frac{\epsilon_1^2}{2}$. Using $\sin\Theta_f=\sin\epsilon_1\simeq\epsilon_1$ and $\frac{\Delta}{2}\sin(2\Theta_f)=-\frac{\Delta}{2}\sin(2\epsilon_1)\simeq-\Delta\epsilon_1$ we get:
        \begin{equation}\label{eq:Linkm1}
          \kappa\,\epsilon_1 \simeq Y(1)=\sqrt{u+\Delta^2\epsilon_1^2}-\Delta\epsilon_1=\sqrt{u}\left(\sqrt{1+c_1^2}-c_1\right)=\sqrt{u}\,e^{-\operatorname{arsinh}c_1} \Longrightarrow \kappa=\frac{e^{-\operatorname{arsinh}c_1}}{c_1}\Delta
        \end{equation}
        where we used $c_1=\epsilon_1 \Delta/\sqrt{u}$.  Therefore in this scaling window $\kappa$ is again linear in $\Delta$, and $\kappa= \hat \kappa \Delta$ with $\hat \kappa=e^{-\operatorname{arsinh}c_1}/c_1.$ Using \eqref{eq:u_window_qm1} we deduce
        \begin{equation}
          c_1=\sqrt{\frac{1}{\hat \kappa (2+ \hat \kappa)}} \quad \Longrightarrow \epsilon_1= 2 e^{-\Delta/4}\hat \kappa^{1/4} c_1^{5/4}= \frac{2 e^{-\Delta/4}}{  \hat\kappa^{3/8} (2+\hat \kappa)^{5/8} },
        \end{equation}
        which gives the scaling window
        \begin{equation}
          (1+q)= \frac{\epsilon_1^2}{2}= \frac{2 e^{-\Delta/2}}{ \hat \kappa^{3/4} (2+\hat \kappa)^{5/4} }, \quad \quad u=4\Delta^2\,e^{-\Delta/2}\,\left(\frac{\hat \kappa}{2 + \hat \kappa} \right)^{\frac{1}{4}}.
        \end{equation}
        This scaling matches again with the regimes analyzed above: the limit $\hat\kappa\to 0$  gives $\hat \kappa= e^{-2\Delta/3}/[2^{1/3} (1+q)^{4/3}]$, consistently with the $q \to -1$ limit of \eqref{eq:AA}.  The scaling variable $\hat \kappa$ diverges as $q\to -1$ as $\hat \kappa \sim \sqrt{2} e^{-\Delta/4}(1+q)^{-1/2}$, consistently with what is obtained plugging $q=-1$ in \eqref{eq:ToExp}. \\
        The tilt grows back to order $\Delta$ in this exponentially small window $1+q\lesssim e^{-\Delta/2}$ near the opposite minimum because the dynamics slows down near the fixed point, and thus the control needs to act to end at the target overlap at a prescribed time. The divergence at $q=-1$ is due to the fact that the tilt couples only to the orthogonal projection of the final configuration ${\bm \sigma}(\hat T)$ to ${\bm \tau}$, which vanishes for ${\bm \sigma}(\hat T)=-{\bm \tau}$.
\end{itemize}

\subsection{Biased ensemble and Legendre transform when $\Delta \to \infty$}\label{sec:Asymptoticpinning}
\begin{figure*}[t]
  \centering
  \includegraphics[width=0.47\textwidth]{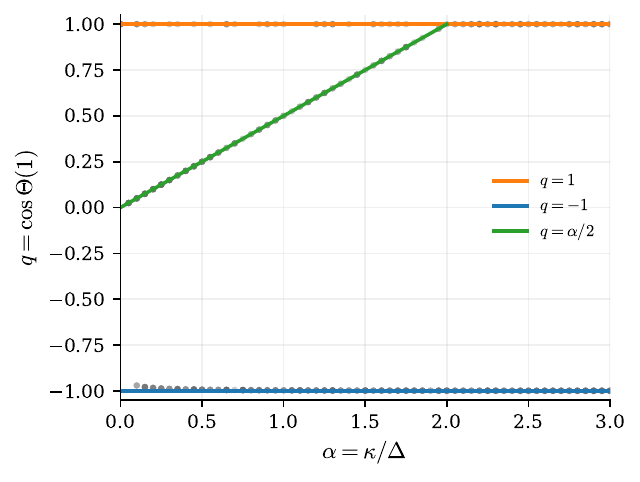}
  \hfill
  \includegraphics[width=0.47\textwidth]{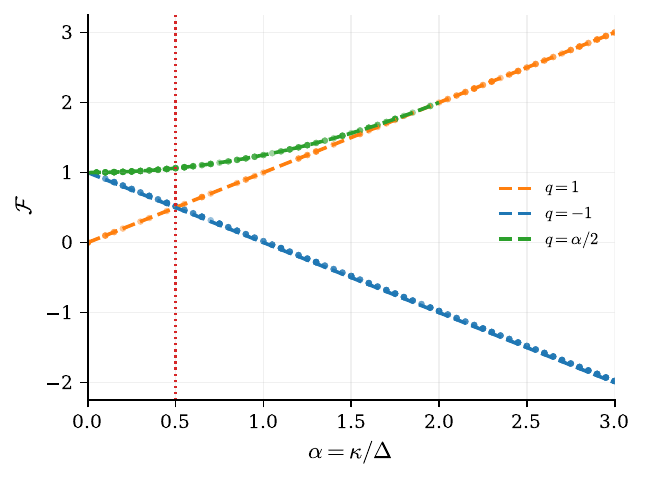}
  \caption{\textit{Left.} Final overlap obtained from solving the FB equations for $(Y,\Theta)$ from independent random initializations,  at $\Delta=10$ and various $\kappa$. The solid curves show the three asymptotic branches $q=1$, $q=-1$, and $q=\alpha/2$.
    \textit{Right.} Corresponding rescaled cost functions $\mathcal{F}$. The dashed lines show the predictions
    $\mathcal F_+=\alpha$,
    $\mathcal F_-=1-\alpha$, and
    $\mathcal F_{\mathrm{int}}=1+\alpha^2/4$.
    The dotted vertical line marks
    $\alpha_c=1/2$.}
  \label{fig:alpha_branches}
\end{figure*}

The more refined scaling analysis shows that for large $\Delta$, $\kappa \sim O(\Delta)$ in three regimes: for $q=1$, for $q \in (0,1)$ and for an exponentially small window around $q=-1$.
We set $\kappa=\alpha \Delta$. A fixed $\alpha \sim O(1)$ thus can correspond to different solutions for $q$, see Fig. \ref{fig:alpha_branches} \emph{Left}. One has to identify which is selected by the Legendre transform. To see this, we consider the cost function \eqref{eq:Cost} evaluated at the optimal control solution, properly rescaled:
\begin{equation}\label{eq:Asympq}
  \frac{1}{\delta^{(N)}} \, \mathcal{C}_{\hat T, \tilde \nu}[{\bm \pi}^*; {\bm \tau}]=\frac{1}{\delta^{(N)}}  \tilde{v}_{\bm \tau}(q, \hat T)+ \frac{\tilde \nu}{\delta^{(N)}}  q \stackrel{\Delta \to \infty}{\longrightarrow} \mathcal{F}(\alpha, q):=  \begin{cases}
    1-q^2+ \alpha  q \quad \quad & 0 <q\leq 1 \\
    1 + \alpha q \quad \quad     & -1<q<0.
  \end{cases} \quad \quad \alpha= \frac{\kappa}{\Delta}.
\end{equation}
As shown in Sec.~\ref{sec:BiasScaling}, a fixed $\alpha$ corresponds to three different values of $q=q(\alpha)$: either $q(\alpha)=\alpha/2$ from \eqref{eq:kappaIn1}, or $q(\alpha)=-1$ from \eqref{eq:Linkm1}, or $q=1$. The latter is a solution of the control problem for any value of $\kappa$, and it is the trivial one (corresponding to a control field that is exactly zero). Plugging these values of $q=q(\alpha)$ in \eqref{eq:Asympq} we get three values of the cost function:
\begin{equation}
  \mathcal{F}\left(\alpha, q(\alpha) \right)=  \begin{cases}
    \mathcal F_+= \alpha\quad \quad                           & [q= 1]   \\
    \mathcal F_-= 1-\alpha\quad \quad                         & [q= -1]  \\
    \mathcal F_{\mathrm{int}}=1+\frac{\alpha^2}{4}\quad \quad & [0<q<1].
  \end{cases}
\end{equation}
Fig.~\ref{fig:alpha_branches} \emph{Right.} shows a comparison of these three costs, together with results for finite $\Delta=10$. One sees that the branch corresponding to $q \in (0,1)$ is always metastable, and therefore not selected by the Legendre transform. This is consistent with the fact that the asymptotic value of the rate function $\tilde{v}_{\bm \tau}(q, \hat T)$ is concave in $q \in (0,1)$, and with the fact that the Legendre transform method only recovers the convex hull of the function. The figure also shows that a first order transition occurs at $\alpha_c=(\kappa/\Delta)_c=1/2$: any $\alpha<1/2$ gives the trivial solution, while arbitrary $\alpha>1/2$ give trajectories with endpoint asymptotically close to $-1$. This is the $\Delta \to \infty$ limit of the pinning transition discussed in Sec.~\ref{sec:pinning}. Notice that Fig. \ref{fig:kappa_crossover} shows that already for $\Delta \simeq 10$, only the $q=1$ and $q=-1$ trajectories are recovered. The Legendre transform method does not allow to get trajectories with endpoint $q \in (-1, 0)$: this would require exponential precision in the choice of the bias $\tilde \nu$, as shown in \eqref{eq:AA}. In this problem the tilted ensemble (meaning, solving the problem at fixed $\tilde{\nu}$ and then choosing the tilt that produces the desired overlap) works only for $\Delta$ relatively small, and fails at $\Delta \to \infty$. The way to proceed here is to enforce directly $q$ as a hard constraint, as it is done within the optimal control formalism.

\end{document}